\documentclass[a4paper,11pt]{article}

\usepackage{authblk}
\usepackage[utf8]{inputenc} 
\usepackage[ruled,lined,algo2e]{algorithm2e}
\usepackage{amssymb}
\usepackage{graphicx}
\usepackage{multirow}

\usepackage{comment}

\usepackage{subfigure}
\usepackage{adjustbox}

\newsavebox{\myfig}

\usepackage[backend=biber,style=authoryear,url=true, isbn=false,doi=false, natbib]{biblatex}
\begin{document}

\title{Nonlinear negotiation approaches for complex-network optimization: a study inspired by Wi-Fi channel assignment \thanks{This is a pre-print of an article published in Group Decision and Negotiation. The final version is available online at \url{https://doi.org/10.1007/s10726-018-9600-z}}
}

\author[1]{Ivan Marsa-Maestre}

\author[1]{Jose Manuel Gimenez-Guzman}

\author[1]{Enrique de la Hoz}

\author[2]{David Orden}

\author[3]{Mark Klein}

\affil[1]{Departamento de Autom\'atica, Universidad de Alcal\'a, Spain.
\texttt{ivan.marsa@uah.es}, \texttt{josem.gimenez@uah.es}, \texttt{enrique.delahoz@uah.es}}

\affil[2]{Departamento de F\'{\i}sica y Matem\'aticas, Universidad de Alcal\'a, Spain.
\texttt{david.orden@uah.es}}

\affil[3]{Center for Collective Intelligence, MIT, USA. \texttt{m\_klein@mit.edu}}

\date{}
\maketitle






\begin{abstract}
At the present time, Wi-Fi networks are everywhere. They operate in unlicensed radio-frequency spectrum bands (divided in channels), which are highly congested. The purpose of this paper is to tackle the problem of channel assignment in Wi-Fi networks. To this end, we have modeled the networks as multilayer graphs, in a way that frequency channel assignment becomes a graph coloring problem. For a high number and variety of scenarios, we have solved the problem with two different automated negotiation techniques: a hill-climber and a simulated annealer. As an upper bound reference for the performance of these two techniques, we have also solved the problem using a particle swarm optimizer. Results show that the annealer negotiator behaves as the best choice because it is able to obtain even better results than the particle swarm optimizer in the most complex scenarios under study, with running times one order of magnitude below. Finally, we  study how different properties of the network layout affect to the performance gain that the annealer is able to obtain with respect to the particle swarm optimizer.

\end{abstract}

\section{Introduction}
\label{intro}

In our current society everything is interconnected, with the Internet network as the prime example. Internet ubiquity and popularity have grown impressively in the last decades. Because of this, many of the current research problems can be modeled as interconnected nodes, i.e. as networks. We can find real-world problems that can be modeled by networks in key strategic fields like transportation~\citep{Ghavidelsyooki17}, energy~\citep{Valori16}, industrial processes~\citep{Bernini16}, medical disciplines like neurology~\citep{Fornito16} and communication networks. This last domain is the one this paper focuses on.

More specifically, we are focused in communication networks where nodes have an important feature: they are self-interested. We will compare two different families of methods to address this kind of problems: optimization techniques and automated negotiation. The first family, optimization techniques, are well suited to large-scale problems, as networked systems use to be. However, these techniques fail when there are self-interested nodes that ignore the optimal solution taking a decision that improves their performance or utility, but severely decreases the total network performance. Thus, there is a number of works that are focused on detecting self-interested nodes \citep{Kumar2016, Banchs2016}. The second family, automated negotiation techniques, can reach solutions in a timely manner and consider selfish nodes in their intrinsic behavior \citep{Ren2009} so nodes will be less predisposed to deviate from the solutions given (in fact, they are usually called \textit{agreements}).
Although automated negotiation techniques could be worth to solve the type of problems we deal with in this paper, they have been barely used to solve complex networked problems \citep{Jonge2015}, as their applications have been focused on designing tools for collaboration, e-commerce or decision-making support \citep{Fujita2017}. In this paper we show that automated negotiation also behaves as a very interesting tool to solve complex network optimization problems where there are conflicts of interest.

The specific problem we tackle in this paper is channel assignment in Wireless Local Area Networks (WLAN) operating in infrastructure mode, i.e. consisting in access points (APs) and clients attached to those APs. This is a complex network problem that includes selfish behavior as nodes are only interested in their performance, not in the global performance of the whole network. Thus, the problem consists of assigning the frequency channel to each AP of the network that minimizes interferences and, therefore, maximizes the performance of their clients. As there are interferences between channels and the number of APs to be assigned with a channel is usually much higher than the total number of available channels, this becomes a very complex problem.

This problem fits within other generic well-known research problems like FAP (Frequency Assignment Problem) \citep{Aardal2007, FAP} and the more generic graph coloring problem \citep{Jensen2011book, Tuza2003}, as frequencies can be considered as colors.
Regarding the graph coloring problem we can emphasize the work \citep{Malaguti2010} where there is a survey of the generic vertex coloring problem (VCP), whose objective is to assign a color to each vertex using different colors on adjacent vertices and minimizing the total number of colors required. In addition to this problem, in \citep{Malaguti2010} it is also included a survey on other generalizations of the VCP, like the \textit{Bandwidth Coloring Problem} (BCP), where distance between colors is taken into account forbidding those colorings where distance between two connected vertices is below a certain value. Other works that consider distances between colors are \citep{Griggs2009, Sharp2007, Bodlaender2000}. All these works considering distances between colors, i.e. including hard restrictions in the graph coloring process, are different from our problem because they are focused on minimizing the largest color assigned to the vertices, while in our problem we have a predefined number of colors (the available spectrum band where the technology is able to operate in) and we want to minimize interferences.
The main works in FAP can be found in the survey \citep{Aardal2007}. Although this survey is mainly devoted to channel assignment in cellular networks, military applications or satellite communications, it also includes a brief mapping of the channel assignment problem to WLAN networks. IEEE 802.11 networks (commercially known as Wi-Fi networks) are the most widespread (and, from a practical point of view, unique) WLAN networks. Although these networks can operate in different unlicensed frequency bands, the most widely used is the 2.4~GHz band, that contains 11 possible and partially overlapped channels. As they are partially overlapped, only three frequencies (the first, the sixth and the eleventh channels) do not collide among them. For that reason it is often considered a three-colors problem \citep{Aardal2007}, although in this work we consider the whole set of possible channels.

More closely related to our specific research are the works directly addressing channel assignment in Wi-Fi networks. It is interesting to highlight that the number of works in this field is quite limited if we consider Wi-Fi impact in our daily lives, as we are surrounded by a large and increasing number of Wi-Fi networks. This scarcity is probably due to the high complexity of the problem, being NP-hard, as stated in \citep{Chieochan2010}. It is precisely in \citep{Chieochan2010} where we can find a survey of channel assignment in Wi-Fi networks. It is interesting to emphasize the works \citep{Mishra2005, Mishra2006} as they are probably the most similar works to ours, not only in terms of their scope (channel assignment for Wi-Fi networks) but also in terms of the problem modelling (the Wi-Fi network is modeled as a graph), although we focus on the use of nonlinear negotiation techniques to solve this type of problems. In \citep{McDiarmid2000, Narayanan2002} authors also use graphs for channel assignment, but not specifically in Wi-Fi. Finally, in \citep{Abusubaih2007} authors propose a coordination protocol for dynamic channel assignment in Wi-Fi networks.

In spite of all the above-mentioned works, the objective and contributions of our proposal are different from them. The first contribution is to model the Wi-Fi infrastructure network as a multilayer graph composed by three layers. The second contribution, and probably the most prominent, is to show that nonlinear negotiation techniques are powerful tools to assign channels to APs in comparison to well-known centralized optimization techniques like a particle swarm optimizer. The third contribution consists of analyzing how graph properties, or network layout in terms of topology, contribute to the performance gains that nonlinear negotiation approaches offer.

The rest of this paper is organized as follows. In Section 2 we describe the graph-based problem model. Section 3 describes the negotiation scenario and approach. Section 4 includes the description of the performed experiments and a discussion of the obtained results. Finally, the last section concludes the paper summarizing our main contributions.

\section{System modelling}
\label{modelado}

\subsection{Wi-Fi networks}
\label{architecture}

Wi-Fi technology is the most widespread technology to deploy wireless local area networks. It is based on the family of IEEE 802.11 standards and it operates in unlicensed frequency bands, with the 2.4~GHz frequency band as the most popular. This band is divided into 11 partially overlapped channels \citep{Ng2012}, although this number is dependent on the world region where the network operates. Our focus is to choose the channel where each access point (AP) will operate, and therefore, the channel where each wireless device (WD) will work, as the channel used by a WD is the same than the one used by its associated AP. This type of Wi-Fi architecture, the most widely deployed, is called infrastructure mode, where we have two different types of nodes: APs and WDs/clients, being examples in this last category devices like laptops, smartphones, TVs... From a user's point of view, the access point is usually a wireless router. Note that WDs are able to communicate to each other only through their associated APs. The decision of the channel used must aim to minimize interferences, and, therefore, optimize network throughput.

\subsection{Multilayer graph model}
\label{graphs}

As it is stated in Section~\ref{intro}, FAP is a specific instance of the more generic graph coloring problem, a widely studied field, some of which main results are summarized in \citep{Jensen2011book, Tuza2003}. For that reason, it is natural to model the problem of assigning channels as a graph. A frequency assignment graph is composed by a set of vertices, that represents the devices, and by edges connecting them. This way, channel assignment is reduced to a vertex coloring problem. With this model and the typical vertex coloring problem (VCP), as described in Section~\ref{intro}, we are not able to capture all the peculiarities of our Wi-Fi channel assignment problem due to the following reasons. First, in Wi-Fi networks we have two different types of vertices, APs and WDs. Therefore, we do not have to color all the vertices, only the ones that represent APs. Note that WDs will take the same color of their AP. We could be tempted to not include WDs in the graph, but this would neglect their effect in terms of interferences. Second, the meaning of the edges in the graph would not be clear, as it could mean interferences or wireless connections or the provider to which an access point belongs to. Third, we do not intend to only avoid monochromatic edges, because, at is has been shown, the distance between colors (or channels) is crucial in Wi-Fi networks in terms of interferences. In conclusion, a plain graph along with the VCP is not accurate enough to capture all the peculiarities of wireless networks \citep{Tragos2013}.


\begin{figure}[tb]
\centerline{
\includegraphics[width=0.7\columnwidth]{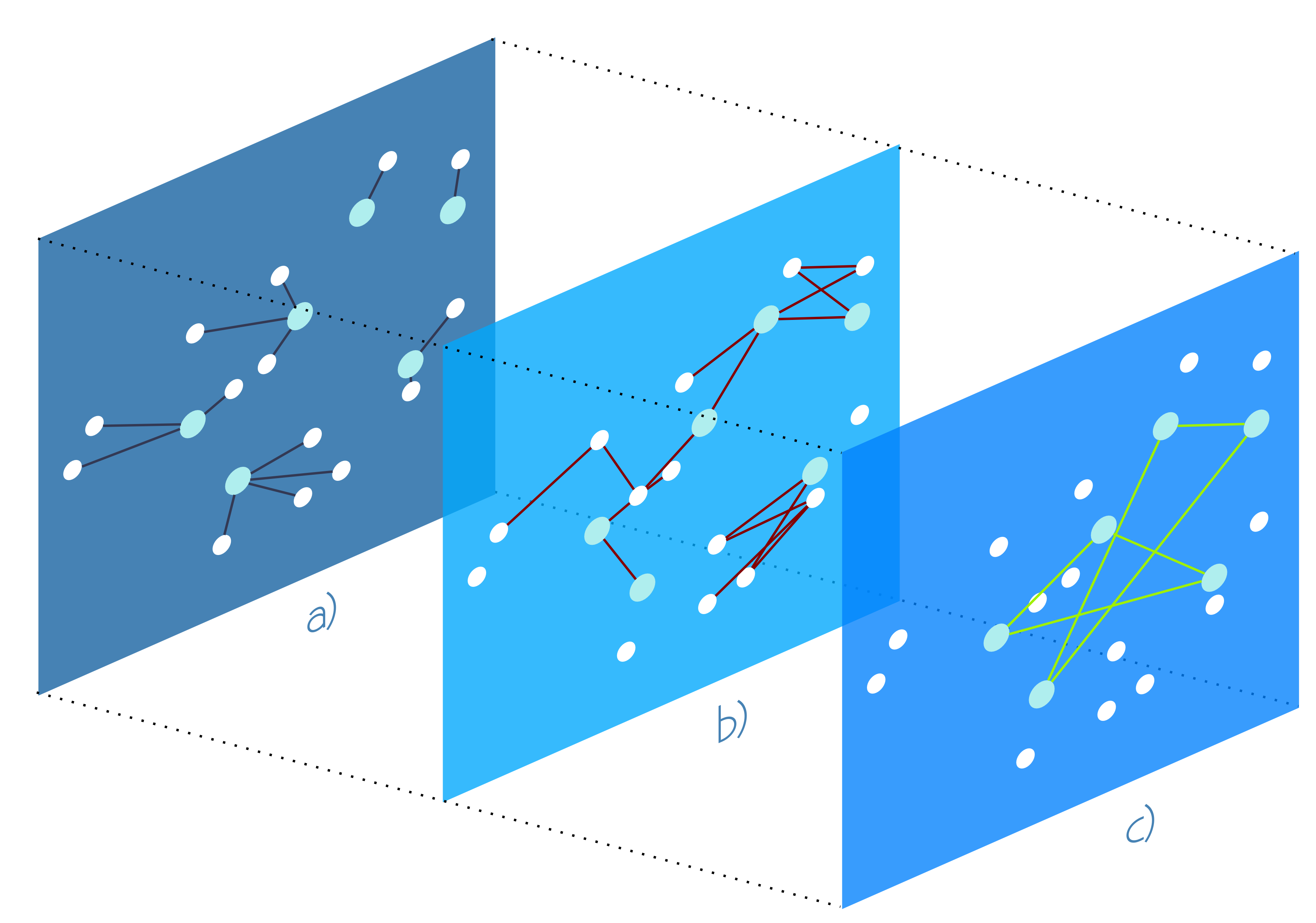}
}
\caption{Multilayer graph model.}
\label{fig:multilayer}
\end{figure}

Once defined the lack of information that a plain graph leads to, we next show the model we propose to more accurately capture the features of Wi-Fi deployments. We use a multilayer network graph \citep{kivela14} in which each layer models a given relation between network elements. More specifically, the graph is composed by three layers, as it can be seen in Fig.~\ref{fig:multilayer}. For each layer we differentiate two different types of vertices: APs and WDs. The information that each layer captures is as follows:

\begin{itemize}
    \item \textit{Layer a} shows the relation between WDs and APs, i.e. in \textit{layer a} edges represent the attachment between each WD and the AP it is associated to. Although a different assumption could be done, we have linked each WD with its closest AP. Remember that when we assign a color to an AP, all the WDs it has associated will also receive the same color, i.e. all the nodes linked in \textit{layer a} will receive the same color.

    \item \textit{Layer b} models the interference between neighboring nodes, so two nodes are linked provided the distance between them is lower than a value $R$, that is obtained from the sensitivity of the reception antennas. Note that two WDs do not collide if they are associated to the same AP, as these communications are coordinated by the AP. The same applies for an AP and all the WDs it has associated. The interference model used is discussed in Section~\ref{propagation}.

    \item Finally, \textit{layer c} links the access points that are controlled by a network provider or Internet Service Provider (ISP). This is a common situation, as it is usual to have a small number of network providers coexisting in the same place. It is very interesting to include this layer, as each provider is able to control their own APs. Of course, this is a general situation, as we could consider that there is only a network provider for the case of the management of the wireless network in a corporation, for example. Moreover, the inclusion of this layer is essential for the negotiation point of view, because an ISP could sacrifice the performance of a node for the sake of improving the performance of others, what enables trade-offs in negotiations.

\end{itemize}

Note that this model does not restrict the communication between WDs associated to different APs, as in the most cases APs are connected to the rest of the network (including the rest of APs and to the Internet in general) through wired links. As these wired links do not affect at all the wireless communications, they do not appear in the multilayer graph.


\subsection{Utility of a channel assignment}
\label{utility}

The performance of a certain channel assignment (network coloring) will be shown in terms of its utility. We define by $U_i$ the utility of node $i$ (either an AP or a WD) as a measure ranging from $0$ to $1$ that depends on the throughput perceived by the user. When the throughput is equal to its maximum value we will have $U_i=1$ and when the throughput decreases the utility will also monotonically decrease. When the node cannot keep connected to the network, and therefore its throughput equals $0$, then $U_i=0$. In other words, the utility of a node can be considered as a normalized throughput. A parallel reasoning can be done in terms of the signal to noise ratio of terminal $i$, defined by $SINR_i$. As it is shown in \citep{Bazzi11}, when $SINR$ is above a certain value $SINR_{max}$, the network throughput reaches its maximum value for a node, so higher $SINR$ values will not lead to better throughputs. Additionally, when $SINR$ is below a certain value $SINR_{min}$ the wireless node cannot keep connected, so its throughput (and utility) equals to $0$. Note that the values for $SINR_{max}$ and $SINR_{min}$ thresholds have been taken from a realistic point of view \citep{Geier}.

For the moment, we have only defined the utility for a certain node. However, as we are interested in the whole utility of a particular channel assignment, we will sum the utility for all the nodes in the graph. Moreover, and for the right operation of the nonlinear negotiation techniques, we have also to define the utility for a provider $p_i$, defined by $U_{p_i}$. This value $U_{p_i}$ is defined as the sum of the utilities for all its APs and the WDs attached to those APs.

\subsection{Interference model}
\label{propagation}

As seen in Section~\ref{utility}, the utility of a particular channel assignment depends on the $SINR$ value of each node ($SINR_i$). As usual, $SINR_i$ is computed as the ratio between the received power of the desired signal and the sum of the received unsought interferences. As APs will have so many $SINR$ values as the number of WDs they have associated, we will take as $SINR$ for each AP the worst case, i.e. the minimum $SINR$ value.

Finally, to be able to compute the signal strengths required for computing $SINR$ we have to take into account interfering signals. In fact, the weight of the edges of the graph \textit{layer b}, as described in Section~\ref{graphs}, represent these interferences. Three are the main elements that affect the signal strength: distance, co-channel interference and activity index. In the following, we describe the impact of these three elements.

Distance is probably the most evident factor that affects signal strength. We have made use of the propagation model defined in \citep{Green2002}. As a result of considering the distance between network nodes, the weight assigned to an edge $ij$ will depend on the distance between nodes $i$ and $j$, so our graph is geometrical and no longer abstract, as it is usually the case for the VCP.

Probably, the most peculiar feature of Wi-Fi networks is the partial overlap between channels. This behavior is captured by adding the co-channel interference index to the signal propagation. In this way, a transmission in channel $i$ will affect another transmission in channel $j$ with a fraction of its nominal power depending on the ``distance'' between channels in the spectrum. The values for this co-channel interference have been taken from the empirical study conducted in \citep{Ng2012}. Note again that the VCP does not consider the distance between colors, so our model is more general.

Finally, we have considered that data flows do not occupy the frequency channels permanently, but they use the spectrum a certain ratio of time. It is obvious that when a node emits with a higher ratio, the interferences it causes to other nodes are more harmful. This fact is represented by the \textit{activity index}.

\section{Wi-Fi frequency assignment as a negotiation process}
\label{negotiation}

As stated above, the problem we want to tackle in this paper is the coordination between APs to select the most appropriate frequencies (channels) in order to reduce interference. We want to address this problem by means of automated negotiation~\citep{Fatima2014}. In this section, we describe the problem as a negotiation process, using the key elements traditionally used to characterize this kind of processes~\citep{Fatima2001}: the negotiation domain, the interaction protocol and the decision mechanisms.

\subsection{Negotiation Domain}

The negotiation domain defines the scope of the negotiation (basically what is negotiated and among whom). In this paper, we propose to see the channel assignment problem as a multiattribute negotiation, where an agreement would be to collectively assign a value to the channel assigned to each of the access points, being these channels the attributes (or \textit{issues}) under negotiation. That is we will consider solutions or contracts $S$ of the form $S=\{s_i|i\in 1,...,n_{AP}\}$, where $n_{AP}$ is the number of access points. Here, $s_i\in\{1,\ldots,11\}$ represents the channel which has been assigned to the corresponding access point $i$.

As stated in the previous sections, we will assume a bilateral negotiation scenario, with two negotiating agents $p_1, p_2$, corresponding to two network providers (commonly ISPs), each of which has jurisdiction over a subset of the APs. This has the advantage that there are more works in the literature to compare with than for the multilateral case (three or more agents). Each agent will have a utility model based on the interference model described in Section 2. With these assumptions, the resulting utility spaces will be non-monotonic and highly rugged, with multiple local optima \citep{Ito2007}.

\subsection{Interaction Protocol}\label{Protocol}

The interaction protocol defines the rules of the negotiation process. There is a wide variety of protocol proposals in the literature for bilateral and multilateral negotiations ~\citep{Rubinstein1982,Ito2007}. Since we expect the utility spaces to be highly rugged, and in a similar way as we did  in~\citep{IJCAI2009}, we will use here a simple text mediation protocol~\citep{Klein2003}. In the following we briefly describe the protocol:
\begin{enumerate}
\item The mediator generates a random candidate contract ($S^c_0$), thus effectively selecting a random channel for each AP as the initial solution. 
\item At time $t$ (starting in $t=0$), the mediator sends contract $S^c_t$ as a proposal for the negotiating agents (i.e., $p_1, p_2$).
\item The negotiating agents then vote on the contract $S^c_t$, either accepting or rejecting it.
\item For $t=t+1$, the mediator builds a new contract $S^c_{t+1}$ taking into account the received feedback and goes back to step~2.
\item After a fixed number of iterations, the process ends and a final agreement is declared.
\end{enumerate}

This protocol definition has to be augmented with the appropriate decision mechanisms (or strategies) for the negotiating agents and the mediator. In the following, we define these decision mechanisms.

\subsection{Decision Mechanisms}
\label{decision}

The decision mechanisms or strategies define how agents behave when facing the different situations that may occur during the negotiation. In our case, we have to define decision mechanisms for the negotiating agents and for the mediator. Negotiating agents have to decide which votes to cast when confronted with a proposal~$S^c$ from the mediator. We have considered two different strategies:

\begin{itemize}
\item \textit{\underline{Hill-climber (HC):}} This is a greedy utility maximization approach, where agents accept a proposal when it yields at least the same utility that the previous mutually accepted proposal.

\item \textit{\underline{Annealer (SA):}} \emph{Simulated annealing}~\citep{Klein2003} is a widely used method to avoid getting stuck in local optima during optimization processes. Contrary to greedy utility maximization, there will be a finite probability $P_a$ for the acceptance of a proposal even when it makes the agent to lose utility. $P_a$ is defined as $P_a=e^\frac{-\Delta u}{\tau}$, where $\Delta u$ is the utility loss for the new proposal, and $\tau$ is an annealing temperature parameter, which linearly decreases to zero during the course of the negotiation. In this way, agents are more flexible at the beginning of the negotiation and become more greedy as the deadline approaches.
\end{itemize}

The mediator, on the other hand, has to decide which new contract to propose to the negotiating agents at each iteration. We have considered here a single-text mediation mechanism \citep{Klein2003} :
\begin{enumerate}
\item The new proposed contract $S^c_{t+1}$ is built from a base contract $S^b$, which is the last contract upon which all negotiating agents have voted \textit{accept}.
\item $S^c_{t+1}$ is obtained by random, single-issue mutation from the base contract $S_{b}$. That is, the mediator randomly varies the assigned frequency for a randomly chosen AP.
\item When the deadline expires, the last contract upon which all negotiating agents have voted \textit{accept} is assumed to be the final agreement.
\end{enumerate}

\section{Scenarios, benchmarks and metrics}
\label{experiments}

\subsection{Considered scenarios}
\label{escenarios}

For the scope of this work, we consider a typical, realistic Wi-Fi configuration, similar to the one used in \citep{Hoz2015}. Readers are referred to this paper for the specifics of the Wi-Fi configuration parameters.

Furthermore, we assume that APs and WDs are static elements. Regarding WDs, we have placed them randomly throughout the scenario, while for the APs we have considered two different types of scenarios. In the first case, APs are distributed randomly. In the second one, APs are located in the junctions of a square grid. According to these positions of the deployed APs, we will call the first type of scenarios \textit{random}, and the second one \textit{square}.

\begin{figure*}[tb]
\centering
\subfigure[Layer a.]{\adjustbox{trim={.3\width} {.18\height} {0.3\width} {.18\height},clip}  {\includegraphics[width=1\textwidth]{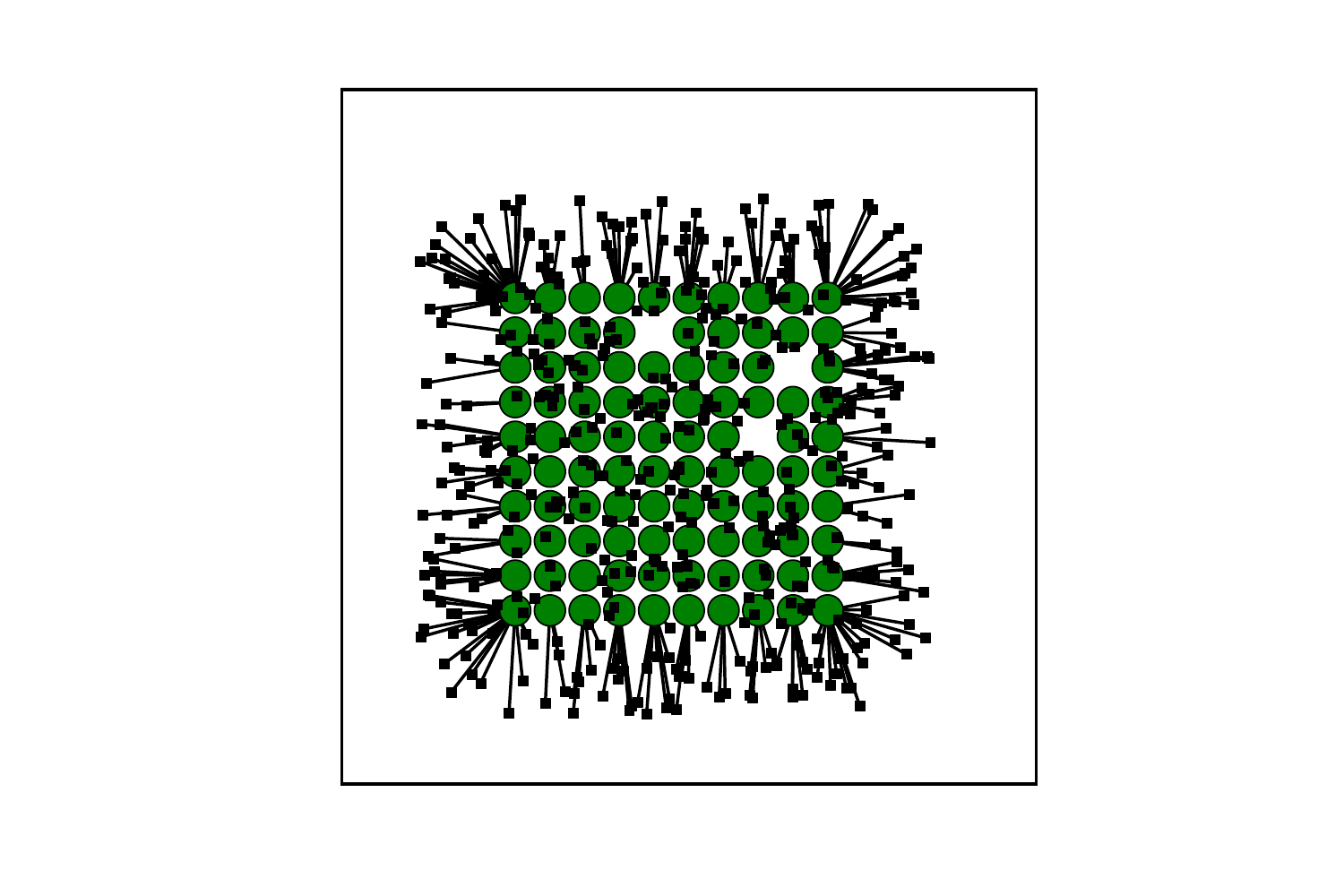}}}\quad
\subfigure[Layer b.]{\adjustbox{trim={.3\width} {.18\height} {0.3\width} {.18\height},clip}  {\includegraphics[width=1\textwidth]{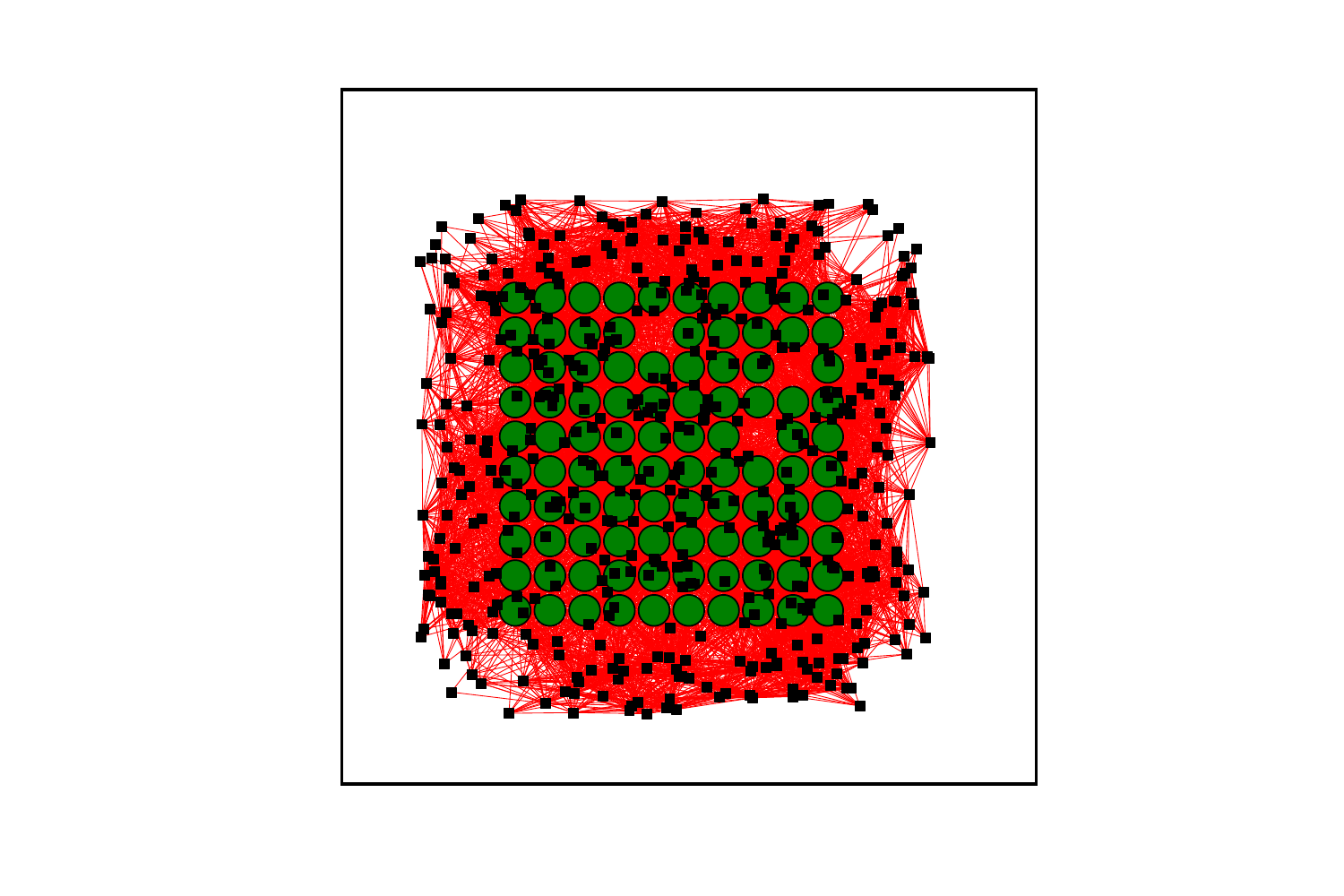}}}
\caption{(100, 500) square scenario example.} \label{fig:graph}
\end{figure*}

With these assumptions, we have generated scenarios varying the number of APs (15, 50 and 100) and the number of clients per AP (1 and 5). We use the nomenclature ($i$, $j$), being $i$ the number of APs and $j$ the number of WDs, having the following combinations: (15, 15), (15, 75), (50, 50), (50, 250), (100, 100) and (100, 500). For each of these combinations of parameters we generated 50 different graphs, for a total of 600 scenarios (300 for random scenarios and 300 for square scenarios). This setting, which is an extended version of the one in \citep{COREDEMA2016}, allowed us to have a wide range of problem sizes (from tens of nodes to roughly one thousand nodes), and also a wide diversity (due to the randomization of node placement). Keep in mind that there is more variability on the number of APs and clients than the one suggested by the parameter set, since we removed from the scenario any AP which had no nearby clients, and vice versa. Finally, for each scenario, we randomly assigned half of the APs to each provider. An example of one of the scenarios under study is the one shown in Fig.~\ref{fig:graph}, that corresponds to a square scenario of the category (100, 500).

\subsection{Analysed techniques}
\label{mecanismos}

We have compared our results with two reference approaches:
\begin{itemize}
\item \emph{\underline{Random Reference}}: the simplest possible technique, where each AP is assigned a channel in a random, uniform manner.
\item \emph{\underline{Augmented Lagrangian Particle Swarm Optimization (ALPSO)}}: this is a parallel particle swarm optimizer, which solves nonlinear non-smooth constrained problems using an augmented Lagrange multiplier approach to handle constraints~\citep{Jensen2011}. This provides the perspective of centralized, complete-information optimization.
\end{itemize}

\subsection{Graph metrics for performance evaluation}
\label{metrics}
One of the research questions we intend to give answer in the long-term is what is the influence of the network-related features of a problem on the adequacy of the different techniques to tackle the problem. With this long-tern goal in mind, we have analyzed our set of scenarios according to a set of metrics taken from the graph theory literature. Some of these metrics are directly or indirectly related to graph size, such as the \textit{order} (number of vertices), or the \textit{diameter} (longest distance between nodes)~\citep{Newman10}. The other metrics give us an idea of the structural properties of the graph regardless of its size. The \textit{Wiener index}~\citep{Wiener47}, for instance, assesses graph complexity from distances between nodes, yielding a metric $W(G)=\frac{1}{2}\sum_{i=0}^{|N|}\sum_{j=0}^{|N|}d(n_i,n_j)$, with $d(n_i,n_j)$ being the shortest distance between vertices $n_i$ and $n_j$. \textit{Graph density} is the global relative connectedness of the graph (i.e. number of edges) when compared to a fully-connected graph.  \textit{Clustering coefficient} is a similar metric but with a local meaning, computing the density for each node's local cluster (i.e. its neighbors and itself) and then averaged. Finally, from the diverse centrality metrics used to rank the importance of vertices in a graph, we have chosen the betweenness centrality, which gives more importance to those nodes which are part of a bigger proportion of the shortest paths in the graph~\citep{Koschutzki05}.

\section{Experimental Results and Discussion}
\label{results}

In this section, we describe and discuss the results of our experiments. For each of the aforementioned 600 scenarios, we did 10 repetitions with each of the benchmarked techniques, recording the achieved social welfare (sum of utilities for both providers, $U_{p_1}+U_{p_2}$) and the computation time.

\begin{table}[bt]
\centering
\caption{Utility for different techniques in random scenarios.}
\label{tab:utilities_rnd}
\small
\renewcommand{\arraystretch}{1.2}
\setlength{\tabcolsep}{.07cm}
\begin{tabular}{ c  c c  c c  c c  c c }
\hline
\multirow{2}{*}{(APs,WDs)} & \multicolumn{2}{c}{Random}  & \multicolumn{2}{c}{HC} & \multicolumn{2}{c}{SA} & \multicolumn{2}{c}{ALPSO}\\
& avg & std & avg & std & avg & std & avg & std\\
\cline{1-9}
(15, 15) & 12.45 &	1.90 &	15.88 &	0.02 &	15.86 &	0.04 &	15.86 &	0.03\\
(15, 75) & 30.57 &	5.18 &	52.53 &	1.35 &	53.85 &	0.50 &	52.95 &	0.93\\
(50, 50) & 29.17 &	4.15 &	50.40 &	0.89 &	51.08 &	0.52 &	50.06 &	0.98\\
(50, 250)& 60.28 &	9.44 &	125.24&	4.71 &	134.96&	2.34 &	125.51&	3.80\\
(100, 100)&45.37 &	5.48 &	84.90 &	2.39 &	88.33 &	1.52 &	83.53 &	2.25\\
(100, 500)&86.21 &	11.68& 	188.13&	7.93 &	208.23&	4.33 &	191.43& 6.25\\
\hline
\end{tabular}
\end{table}

\begin{table}[bt]
\centering
\caption{Utility for different techniques in square scenarios.}
\label{tab:utilities_sq}
\small
\renewcommand{\arraystretch}{1.2}
\setlength{\tabcolsep}{.07cm}
\begin{tabular}{ c  c c  c c  c c  c c }
\hline
\multirow{2}{*}{(APs,WDs)} & \multicolumn{2}{c}{Random}  & \multicolumn{2}{c}{HC} & \multicolumn{2}{c}{SA} & \multicolumn{2}{c}{ALPSO}\\
& avg & std & avg & std & avg & std & avg & std\\
\cline{1-9}
(15, 15) &16.57&	2.67&	24.18&	0.05&	24.15&	0.06&	24.15&	0.06\\
(15, 75) &42.11&	7.01&	74.59&	1.91&	76.91&	0.69&	74.77&	1.68\\
(50, 50) &32.96&	4.43&	60.26&	1.34&	61.80&	0.82&	59.48&	1.45\\
(50, 250)&68.86&	9.69&	145.28&	5.21&	157.19&	2.63&	144.12&	4.32\\
(100, 100)&49.36&	4.55&	91.05&	2.40&	94.64&	1.63&	89.11&	2.38\\
(100, 500)&102.00&	10.44&	205.90&	7.10&	221.68&	4.20&	201.91&	5.12\\
\hline
\end{tabular}
\end{table}

Firstly, we study the performance of the evaluated techniques in the different scenario categories according to the scenario generation parameters (number of APs and number of clients per AP). Table~\ref{tab:utilities_rnd} shows, for the random scenarios, the average utility obtained by each approach for all the graphs in each category, while Table~\ref{tab:utilities_sq} is its equivalent for the square scenarios. We can see that, for the less complex scenarios, all approaches but \textit{random} perform reasonably well, with a non-significant little advantage for the hill climber (\textit{HC}). As the scenarios grow more complex, we can see the performance of the \textit{random} approach turning worse, which is reasonable since the size of the solution space becomes larger. We can also note significant increasing distance between the performances of the hill climber and the annealer (\textit{SA}) negotiators. This difference in the performance of the two approaches confirms our hypothesis that these scenarios are highly nonlinear \citep{Klein2003,IJCAI2009}, since one of the strengths of the annealer is its ability to escape from local optima. We can also see that, for the more complex scenarios, the \textit{SA} negotiator significantly outperforms the particle swarm optimizer (\textit{ALPSO}). This is a remarkable result, specially taking into account that \textit{SA} reaches the optimum faster than the \textit{ALPSO} optimizer. Tables~\ref{tab:time_rnd} and~\ref{tab:time_sq} show the average computation times for both approaches. We can see that, in the largest scenarios, the SA negotiator is roughly from 8 to 10 times faster than the complete information optimizer. If we compare the utility of the random scenarios and their counterpart square scenarios, we perceive that the utility in random scenarios is lower. This is due to the fact that in the square scenarios, as the APs are evenly distributed, the number of APs that have no nearby clients, and therefore are removed from the graph, is lower than in the random setting.

\begin{table}[bt]
\centering
\caption{Run time (in seconds) for different techniques in random scenarios.}
\label{tab:time_rnd}
\small
\renewcommand{\arraystretch}{1.2}
\setlength{\tabcolsep}{.13cm}
\begin{tabular}{ c  c c  c c  c c }
\hline
\multirow{2}{*}{(APs,WDs)} & \multicolumn{2}{c}{HC} & \multicolumn{2}{c}{SA} & \multicolumn{2}{c}{ALPSO}\\
& avg & std & avg & std & avg & std\\
\cline{1-7}
(15, 15) & 0.53 &	0.21 &	0.64 &	0.22 &	0.25 &	0.19\\
(15, 75) & 5.79 &	1.22 &	5.96 &	1.23 &	5.86 &	2.00\\
(50, 50) & 5.22 &	1.16 &	5.40 &	1.17 &	11.91 &	5.02\\
(50, 250)& 69.39&	6.44 &	69.32&	6.36 &	285.89&	74.37\\
(100, 100)&22.01&	2.96 &	22.15&	2.99 &	108.14&	31.39\\
(100, 500)&330.38&	17.23&	326.90&	16.61&	3225.63&817.93\\
\hline
\end{tabular}
\end{table}

\begin{table}[bt]
\centering
\caption{Run time (in seconds) for different techniques in square scenarios.}
\label{tab:time_sq}
\small
\renewcommand{\arraystretch}{1.2}
\setlength{\tabcolsep}{.13cm}
\begin{tabular}{ c  c c  c c  c c }
\hline
\multirow{2}{*}{(APs,WDs)} & \multicolumn{2}{c}{HC} & \multicolumn{2}{c}{SA} & \multicolumn{2}{c}{ALPSO}\\
& avg & std & avg & std & avg & std\\
\cline{1-7}
(15, 15) &1.35&	0.12&	1.55&	0.13&	0.65&	0.10\\
(15, 75) &17.30&	1.41&	13.64&	0.99&	13.25&	2.81\\
(50, 50) &13.05&	1.07&	13.64&	1.23&	23.53&	4.42\\
(50, 250) &115.12&	4.74&	122.93&	4.54&	362.52&	65.05\\
(100, 100) &47.97&	3.23&	47.02&	3.22&	187.34&	33.32\\
(100, 500) &433.16&	12.02&	463.59&	14.75&	3985.70&797.10\\
\hline
\end{tabular}
\end{table}

\begin{figure*}[p]
\centering
\subfigure[(50, 50).]{\adjustbox{trim={.01\width} {.01\height} {0.01\width} {.01\height},clip}  {\hspace{8mm}\includegraphics[width=0.38\textwidth, height=4.5cm]{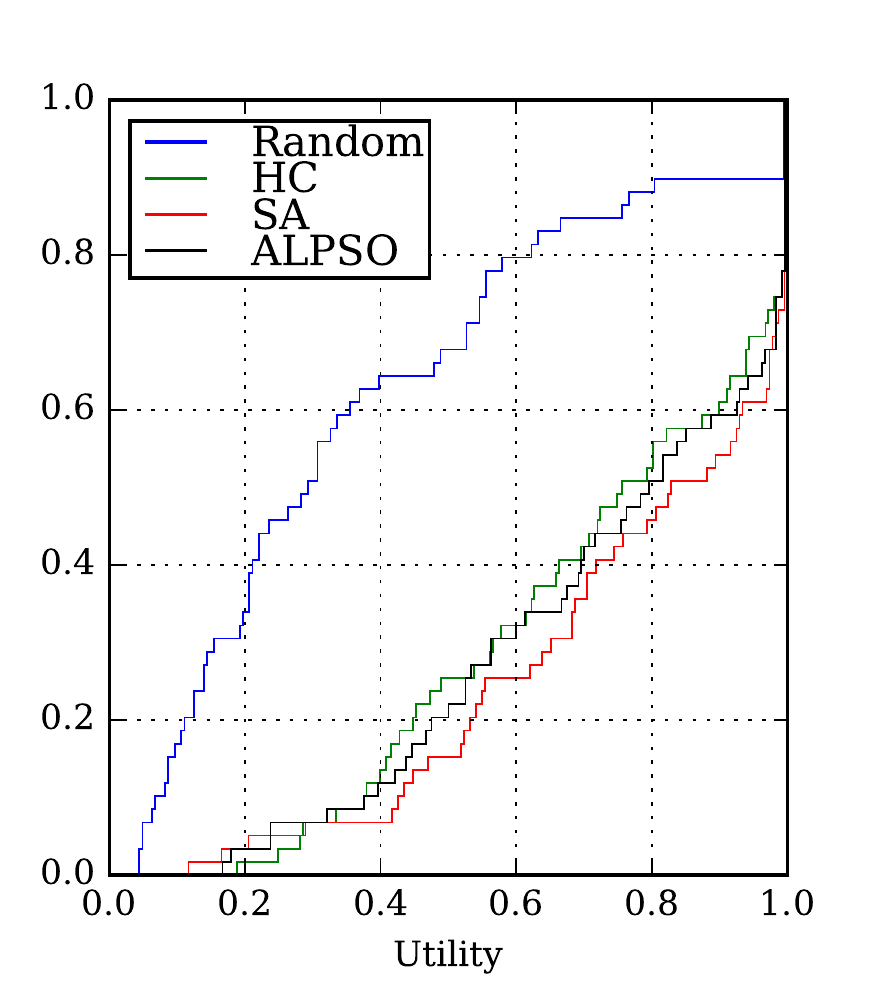}
\hspace{-5mm}
\includegraphics[width=0.55\textwidth, height=4.5cm, trim={2.2cm 0 0 0},clip]{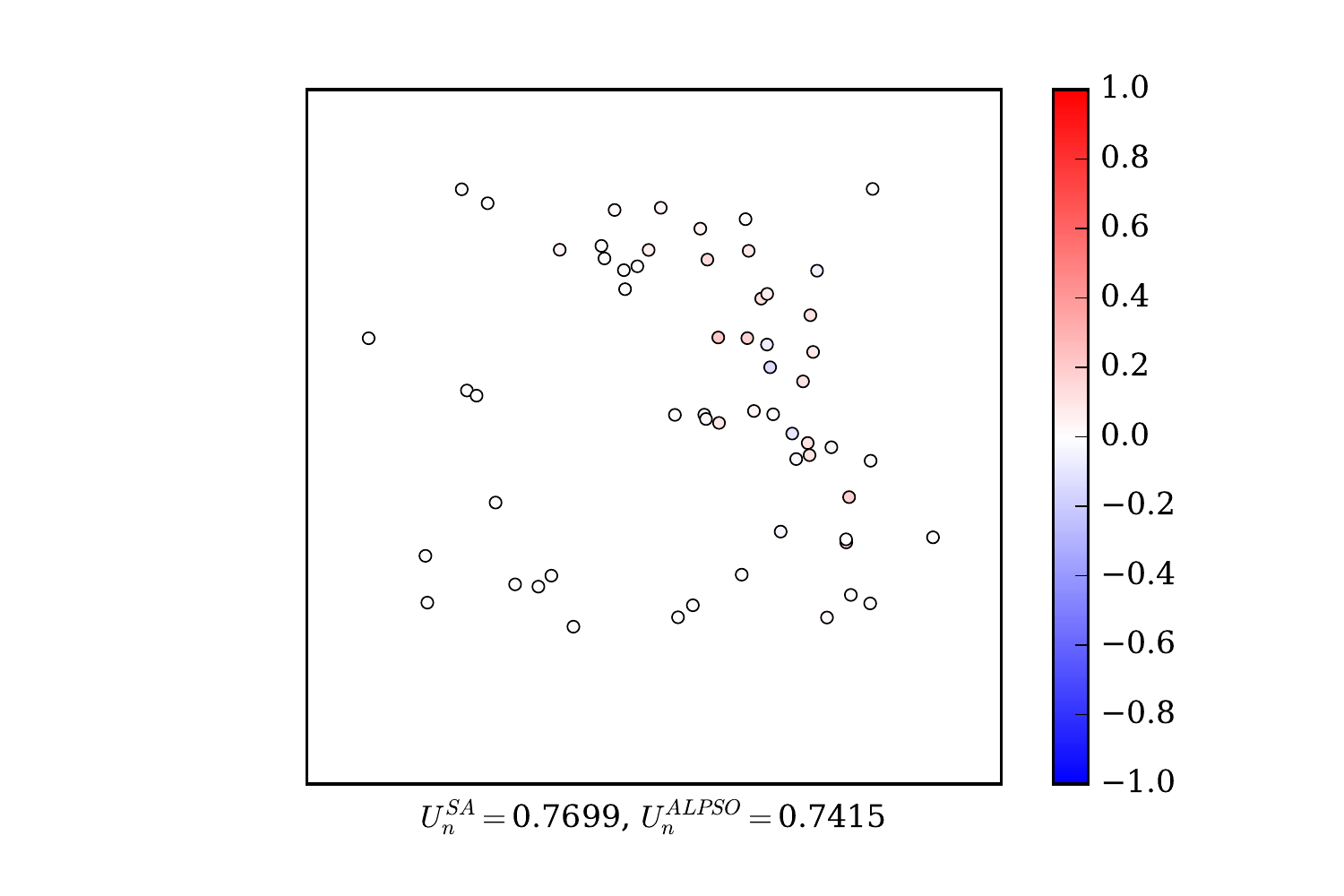}}}\hspace{2mm}
\subfigure[(50, 250).]{\adjustbox{trim={.01\width} {.01\height} {0.01\width} {.01\height},clip}  {\hspace{8mm}\includegraphics[width=0.38\textwidth, height=4.5cm]{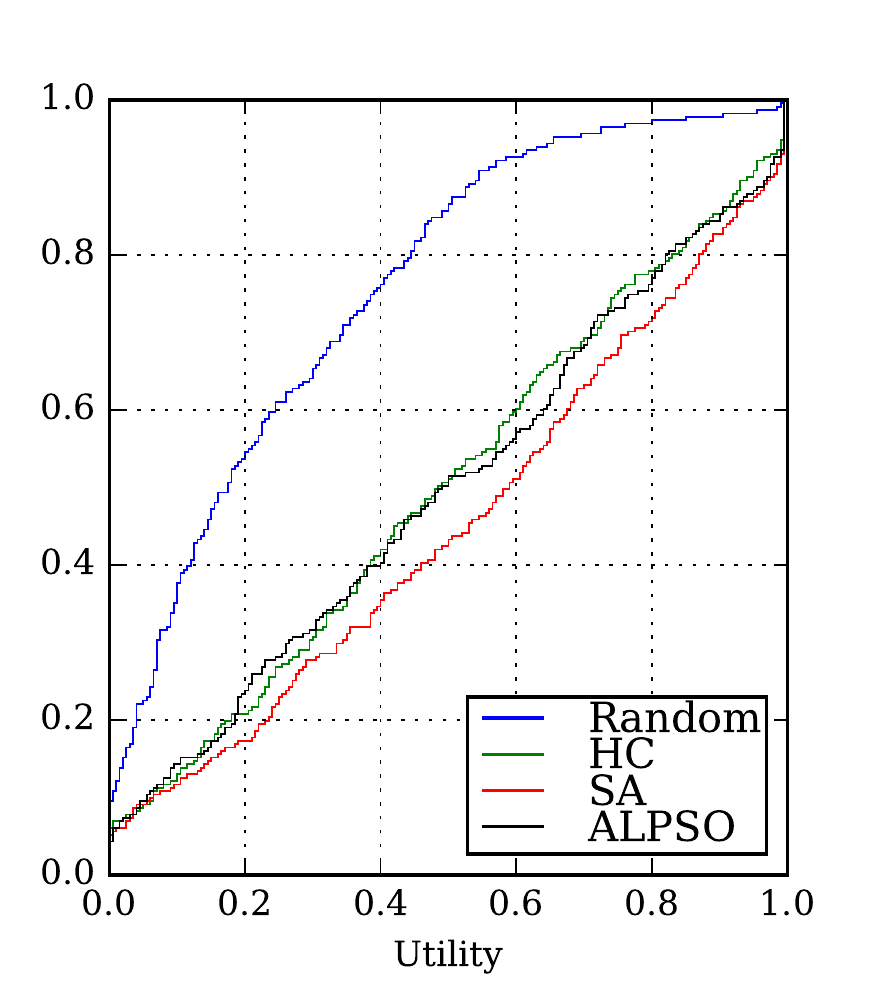}
\hspace{-5mm}
\includegraphics[width=0.55\textwidth, height=4.5cm, trim={2.2cm 0 0 0},clip]{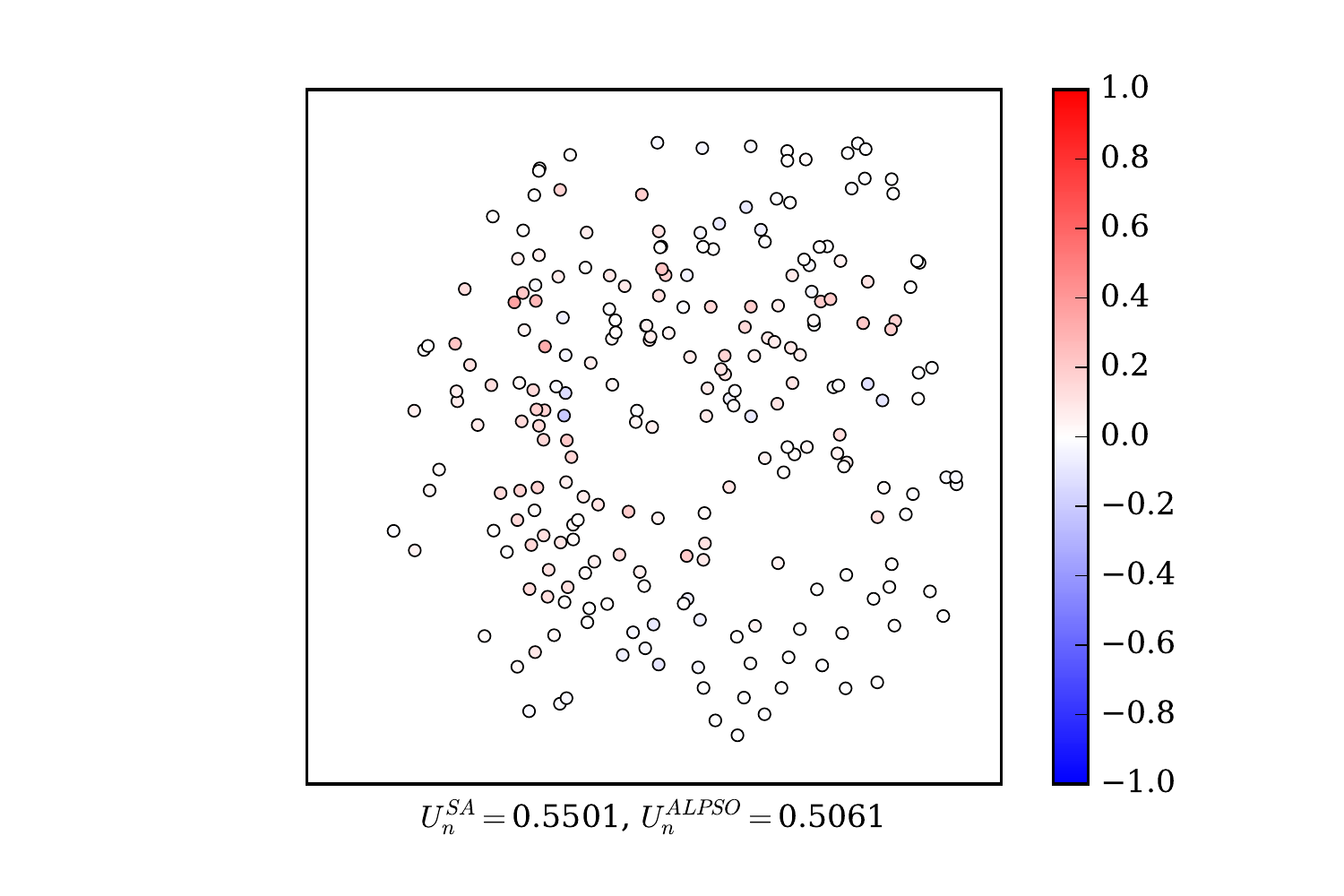}}}\hspace{2mm}
\subfigure[(100, 100).]{\adjustbox{trim={.01\width} {.01\height} {0.01\width} {.01\height},clip}  {\hspace{8mm}\includegraphics[width=0.38\textwidth, height=4.5cm]{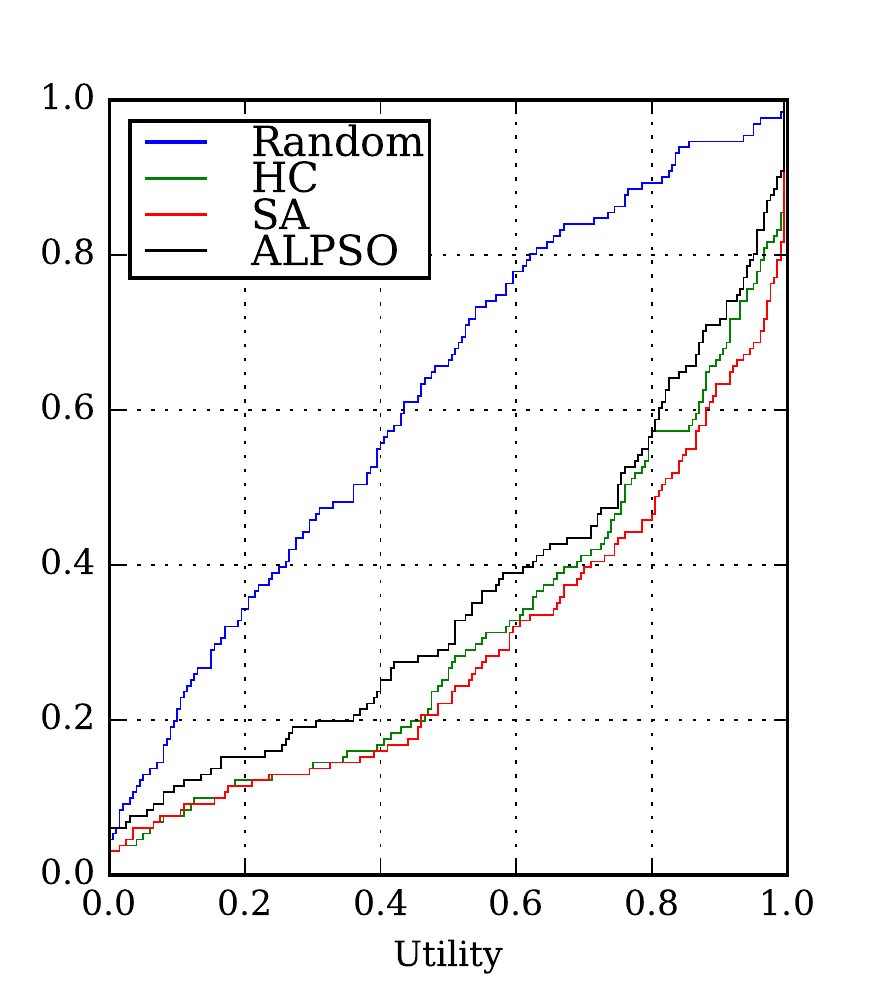}
\hspace{-5mm}
\label{fig:cdf_rnd_100_100}
\includegraphics[width=0.55\textwidth, height=4.5cm, trim={2.2cm 0 0 0},clip]{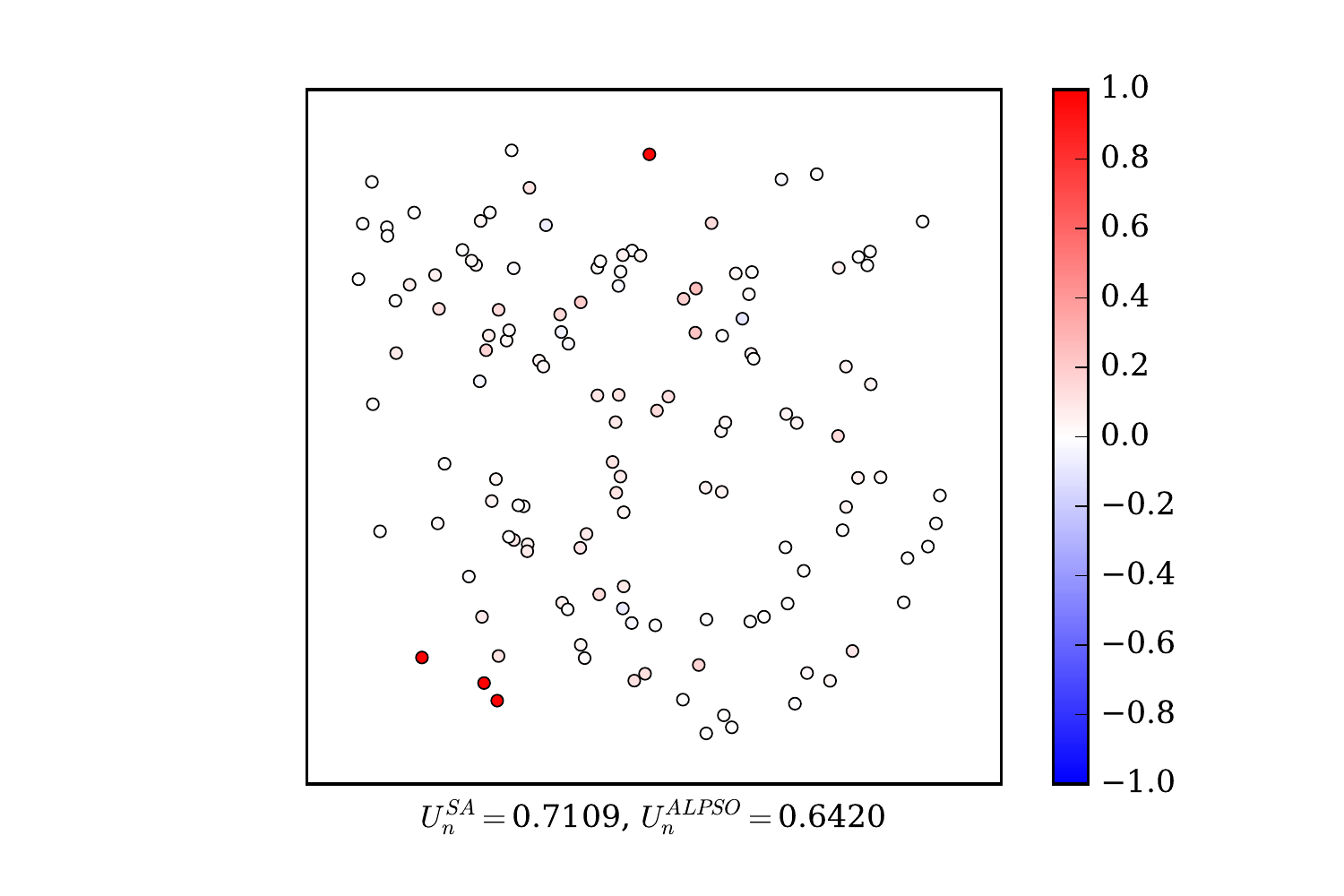}}}\hspace{2mm}
\subfigure[(100, 500).]{\adjustbox{trim={.01\width} {.01\height} {0.01\width} {.01\height},clip}  {\hspace{8mm}\includegraphics[width=0.38\textwidth, height=4.5cm]{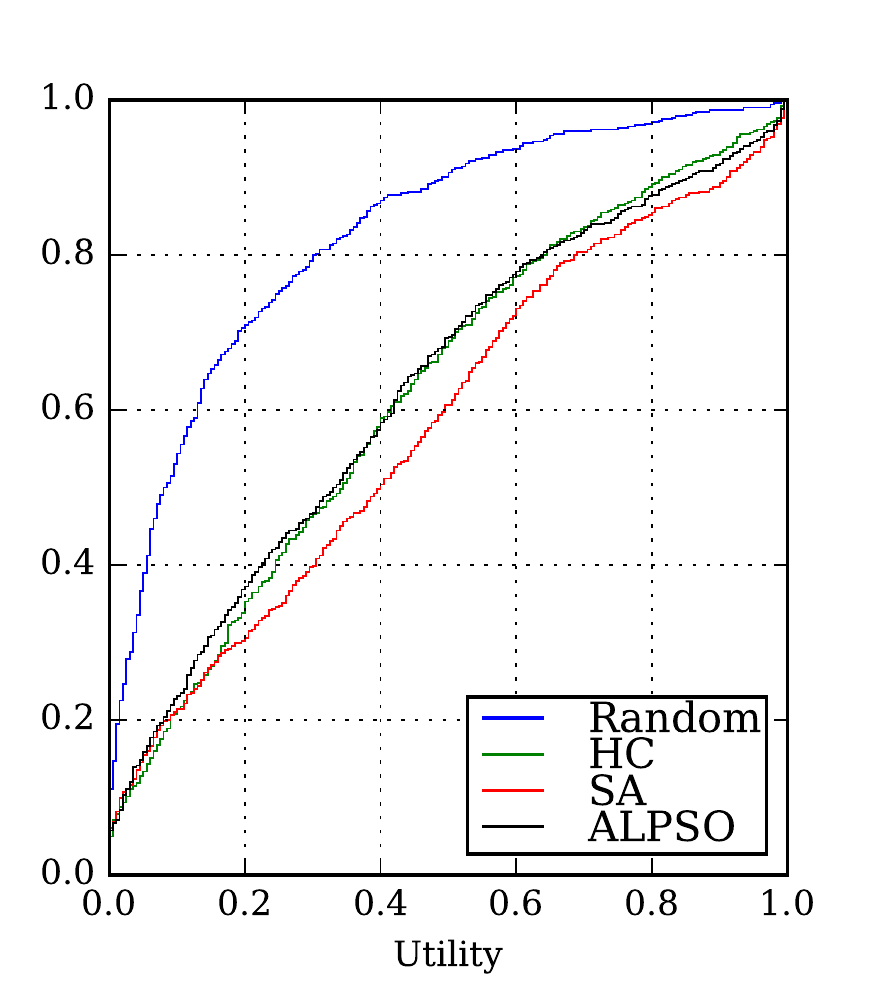}
\hspace{-5mm}
\includegraphics[width=0.55\textwidth, height=4.5cm, trim={2.2cm 0 0 0},clip]{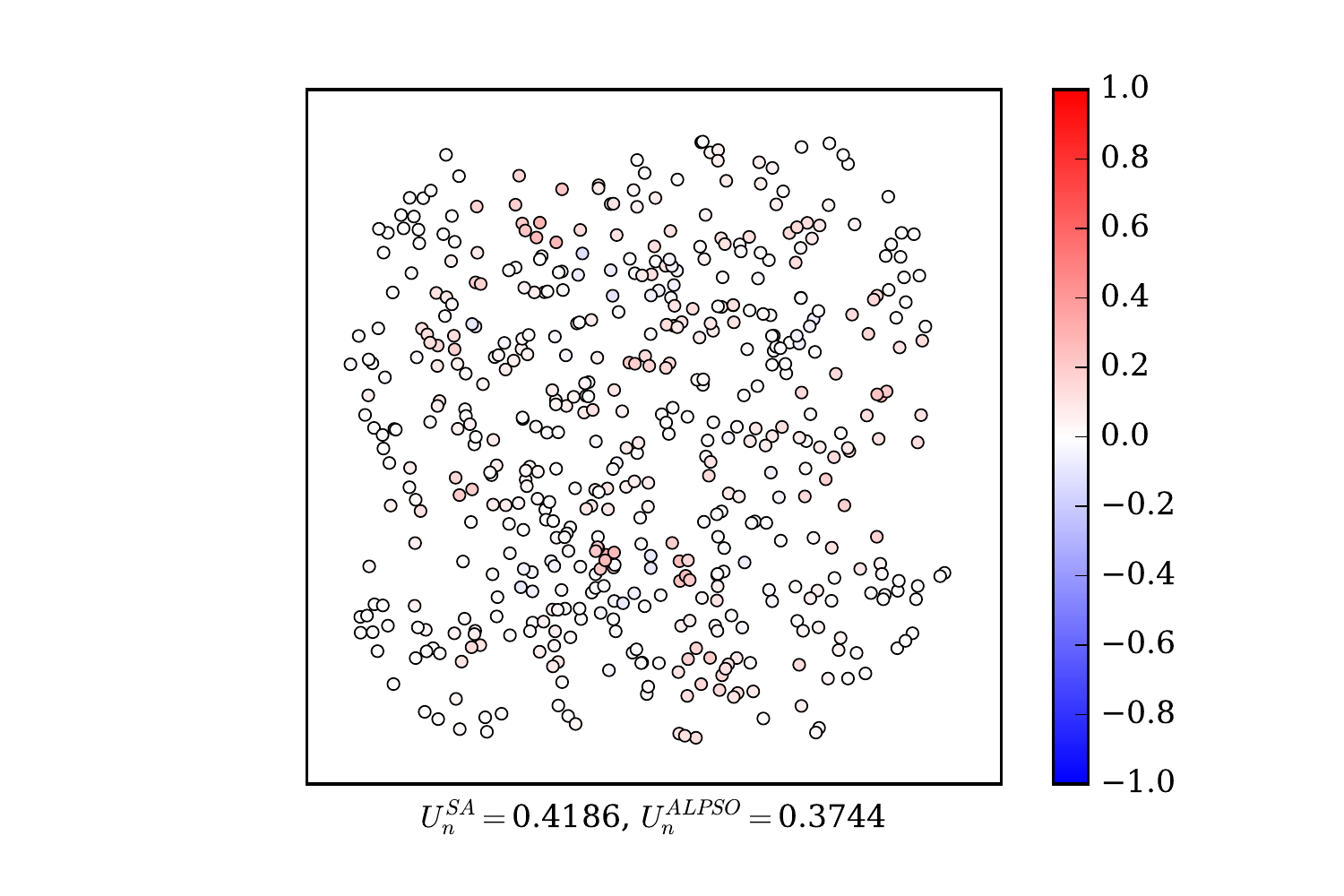}}}\hspace{2mm}
\vspace{-3mm}
\caption{cdf of the utility and comparison of the utility achieved by SA and ALPSO in random scenarios.} \label{fig:cdf_rnd}
\end{figure*}

\begin{figure*}[p]
\centering
\subfigure[(50, 50).]{\adjustbox{trim={.01\width} {.01\height} {0.01\width} {.01\height},clip}  {\hspace{8mm}\includegraphics[width=0.38\textwidth, height=4.5cm]{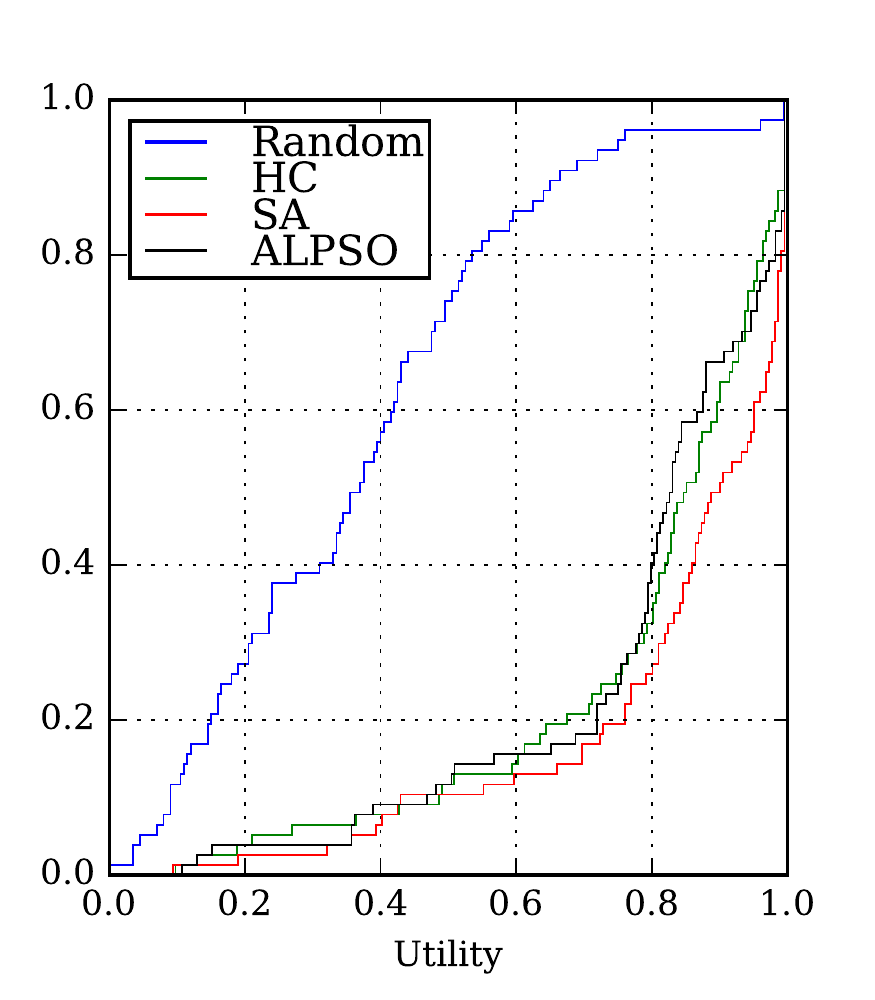}
\hspace{-5mm}
\includegraphics[width=0.55\textwidth, height=4.5cm, trim={2.2cm 0 0 0},clip]{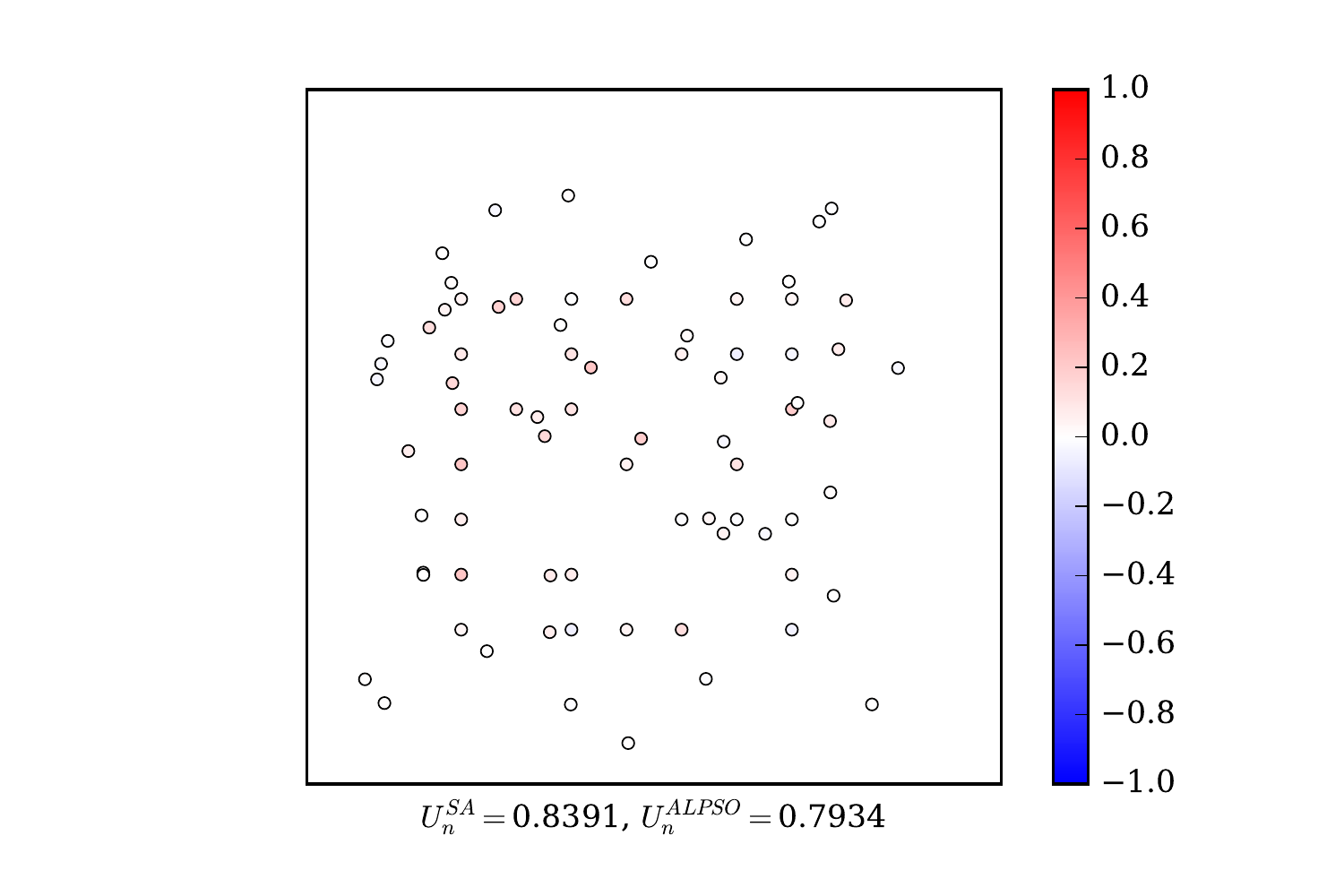}}}\hspace{2mm}
\subfigure[(50, 250).]{\adjustbox{trim={.01\width} {.01\height} {0.01\width} {.01\height},clip}  {\hspace{8mm}\includegraphics[width=0.38\textwidth, height=4.5cm]{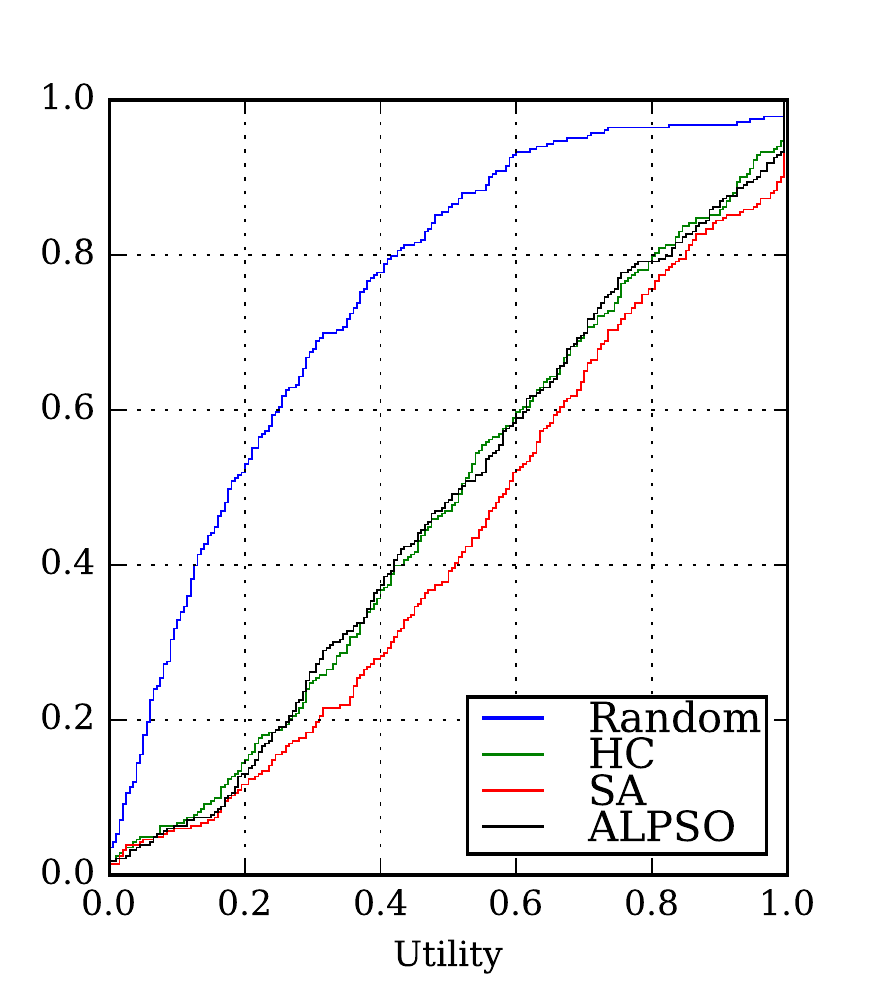}
\hspace{-5mm}
\includegraphics[width=0.55\textwidth, height=4.5cm, trim={2.2cm 0 0 0},clip]{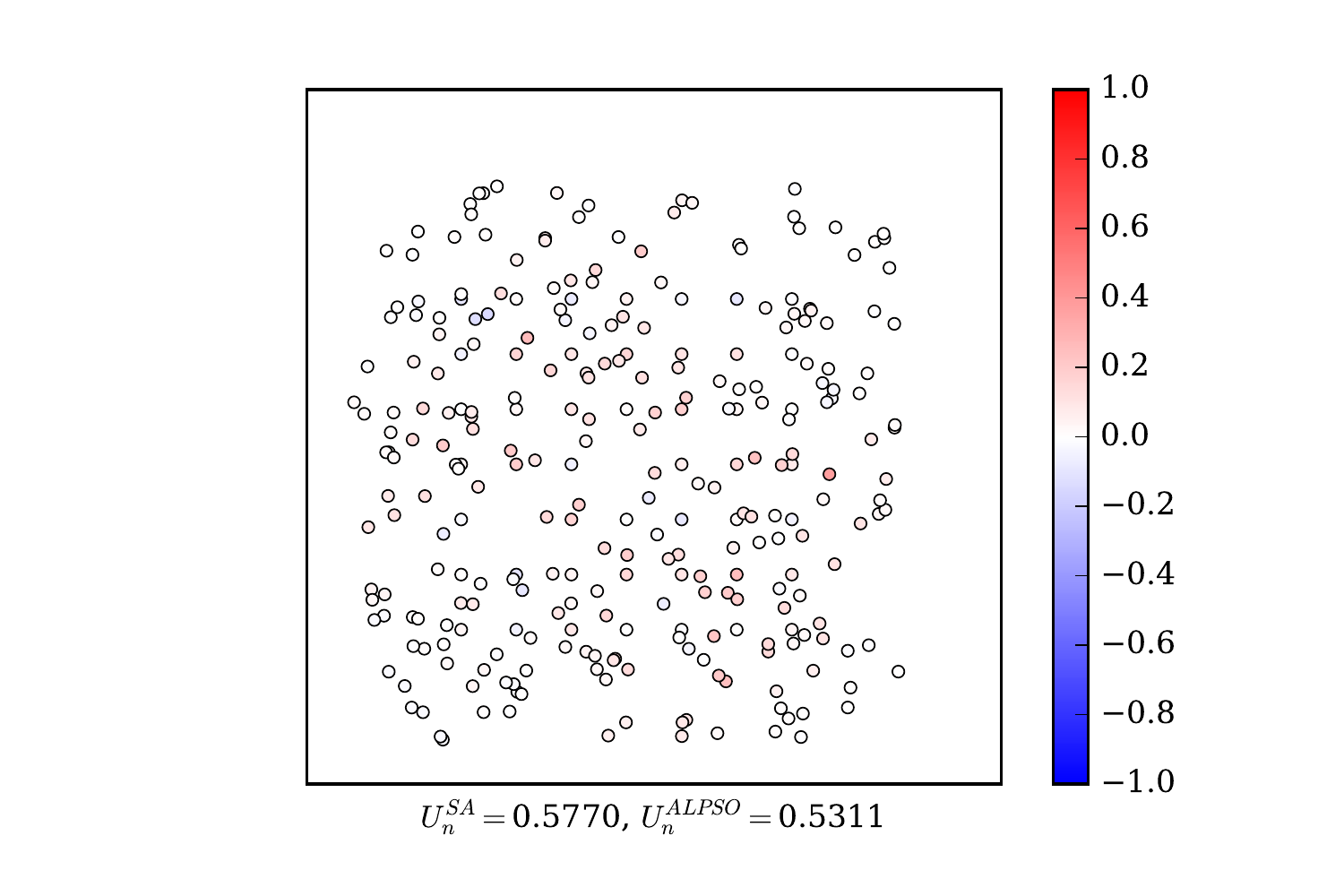}}}\hspace{2mm}
\subfigure[(100, 100).]{\adjustbox{trim={.01\width} {.01\height} {0.01\width} {.01\height},clip}  {\hspace{8mm}\includegraphics[width=0.38\textwidth, height=4.5cm]{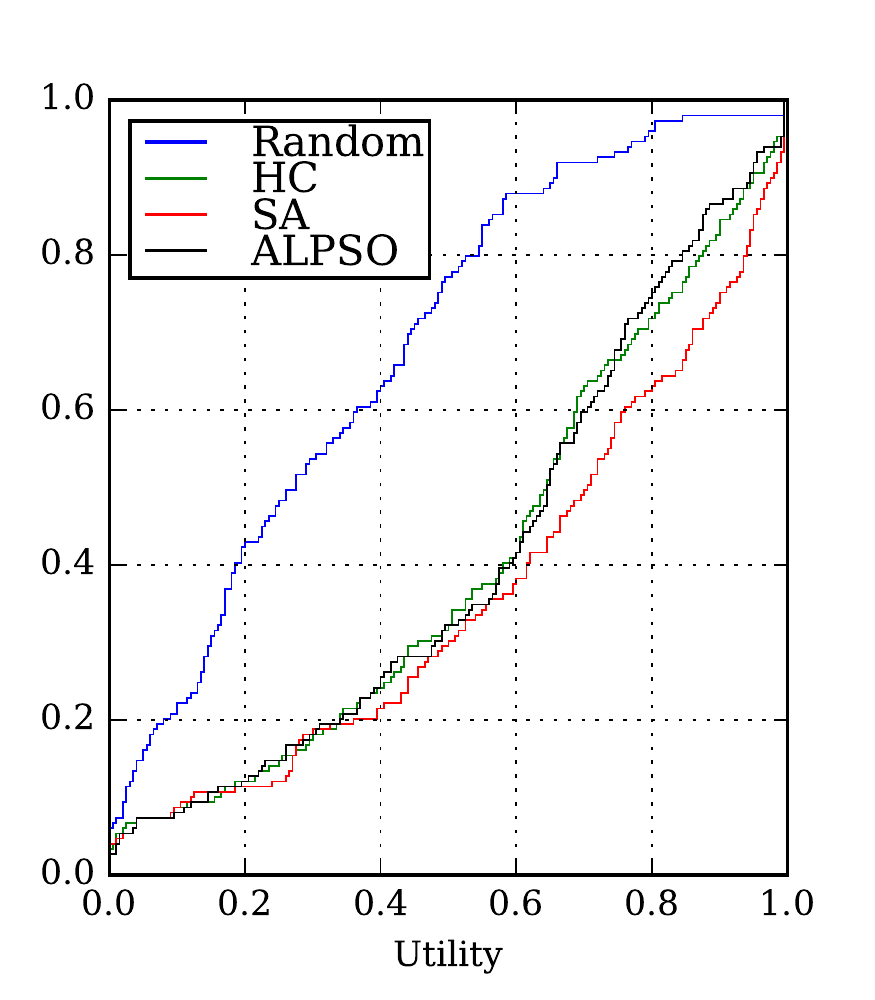}
\hspace{-5mm}
\includegraphics[width=0.55\textwidth, height=4.5cm, trim={2.2cm 0 0 0},clip]{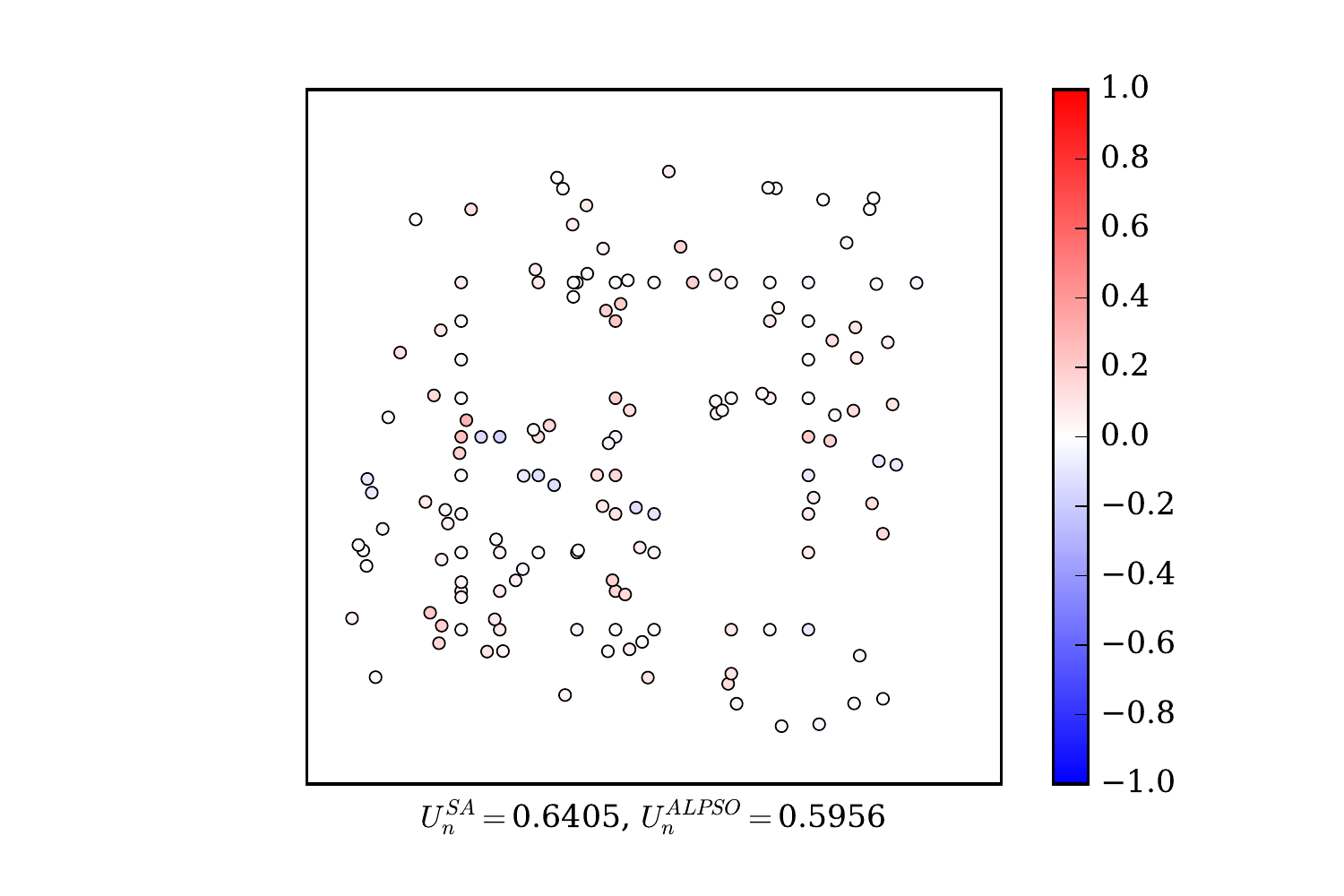}}}\hspace{2mm}
\subfigure[(100, 500).]{\adjustbox{trim={.01\width} {.01\height} {0.01\width} {.01\height},clip}  {\hspace{8mm}\includegraphics[width=0.38\textwidth, height=4.5cm]{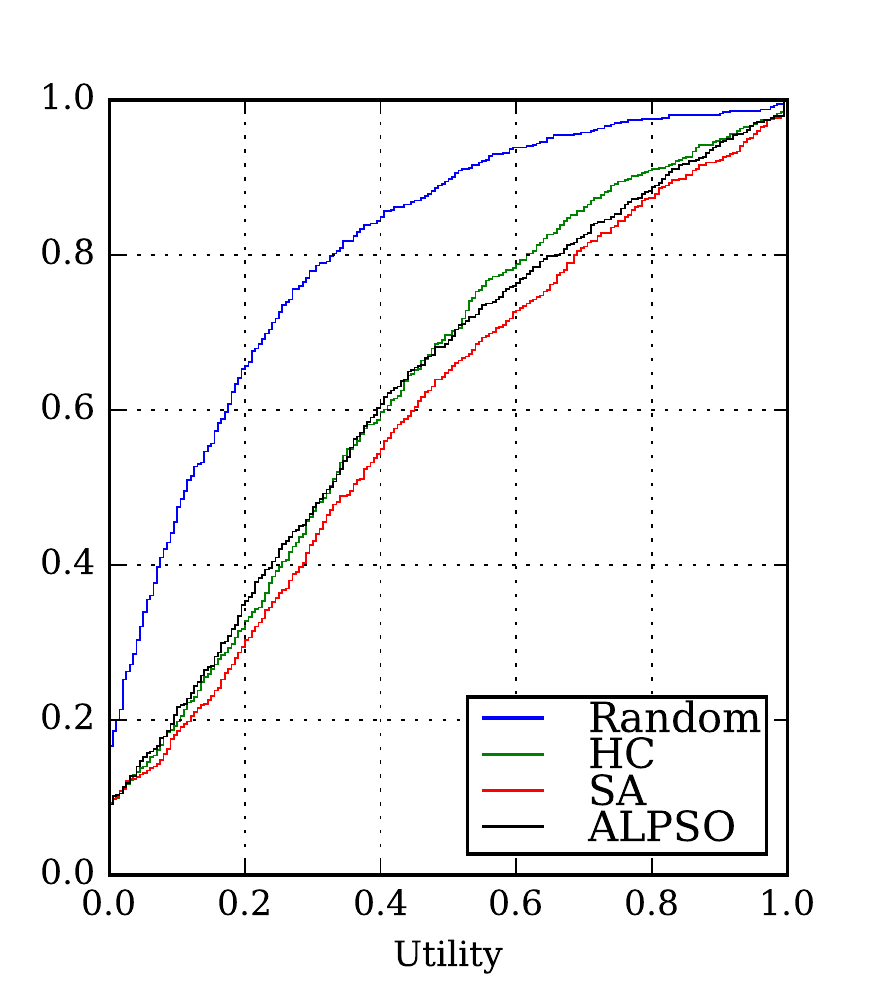}
\hspace{-5mm}
\includegraphics[width=0.55\textwidth, height=4.5cm, trim={2.2cm 0 0 0},clip]{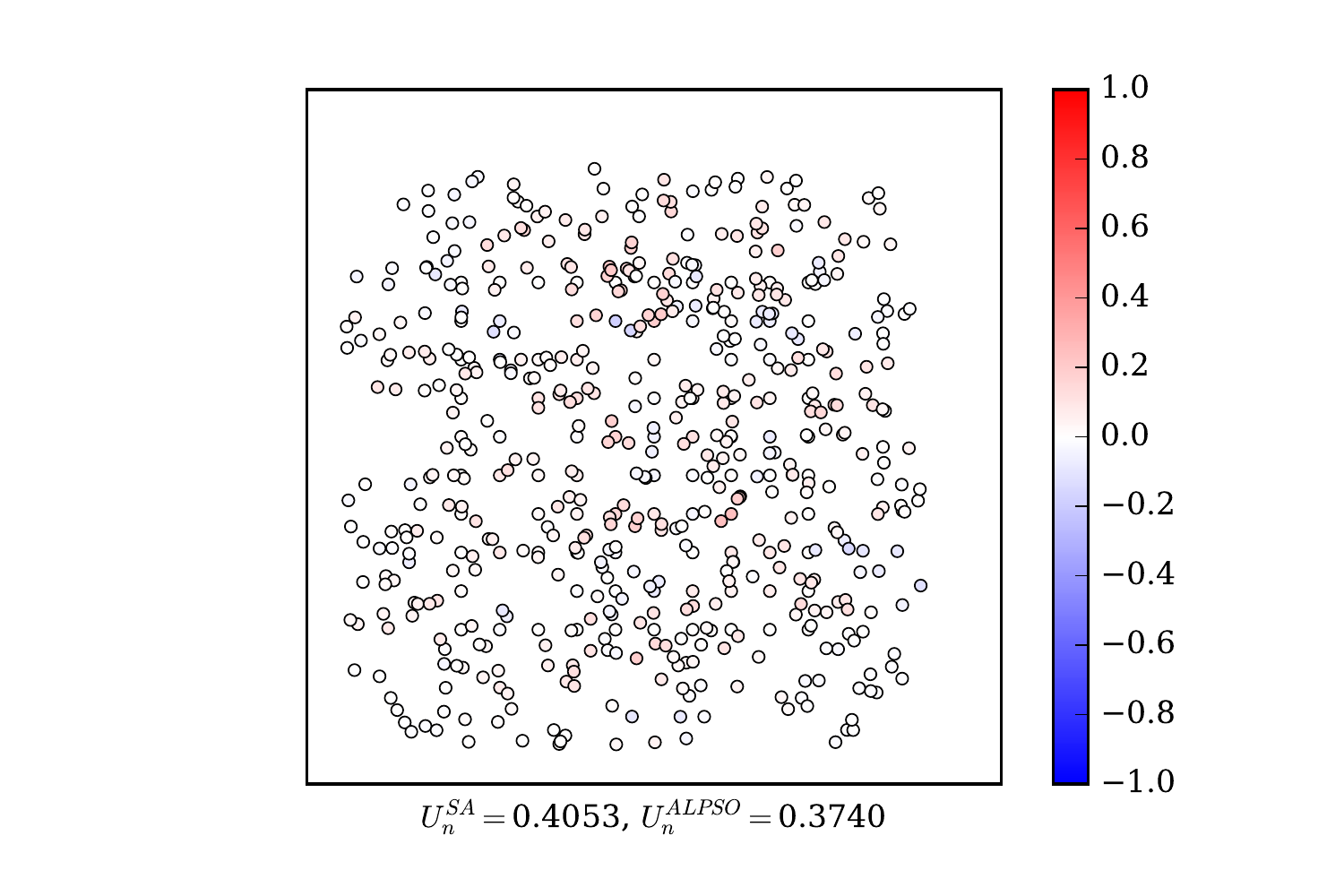}}}\hspace{2mm}
\vspace{-3mm}
\caption{cdf of the utility and comparison of the utility achieved by SA and ALPSO in square scenarios.} \label{fig:cdf_sq}
\end{figure*}

Tables~\ref{tab:utilities_rnd} and~\ref{tab:utilities_sq} offer us a first and clear comparison between the different techniques under study. However, Figs.~\ref{fig:cdf_rnd} and~\ref{fig:cdf_sq} show us a more profound insight for some specific selected scenarios (for the sake of space), although the same conclusions can be given for the rest. Note that, also for the sake of brevity, we have focused on the larger scenarios, omitting the categories (15, 15) and (15, 75). Figures on the left show the cumulative distribution function (cdf) of the utility experienced by the different nodes in the network, so a point ($x$, $y$) represents that a fraction of $y$ nodes have a utility below $x$. For that reason, lower curves are better. In all cases, we can conclude that the worst technique is clearly \textit{random}, while the best one is \textit{SA}. Finally, to compare the performance of \textit{SA} and \textit{ALPSO}, in the figures on the right we show heat maps of the difference in the utility achieved by each node for \textit{SA} and \textit{ALPSO}, i.e. $U_i^{SA}-U_i^{ALPSO}$, $\forall i$, being $i$ the $i$-th node. For that reason, red colorings represent situations where \textit{SA} outperforms \textit{ALPSO}, while blue colorings represent the opposite situation. Note also that each heat map in the right corresponds to the same scenario whose results are shown in the cdf on the left. From those heat maps we can conclude that the utilities given by \textit{SA} and \textit{ALPSO} are not very differently distributed, and that the differences between both techniques are distributed among a high number of nodes, i.e. many of the nodes are light red colored. Note that in some few cases, we have noted the situation given in Fig.~\ref{fig:cdf_rnd_100_100}, where there are three nodes in dark red, so \textit{SA} greatly improves \textit{ALPSO} for them. It is also important to point out that we have not found any point in dark blue for any of the 600 scenarios under study. Finally, heat maps also show in their lower part the mean utility achieved by \textit{SA} ($U_n^{SA}$) and \textit{ALPSO} ($U_n^{ALPSO}$).

\begin{figure*}[tb]
\centering
\subfigure[Order.]{\adjustbox{trim={.02\width} {.07\height} {0.1\width} {.1\height},clip}  {\includegraphics[width=0.54\textwidth]{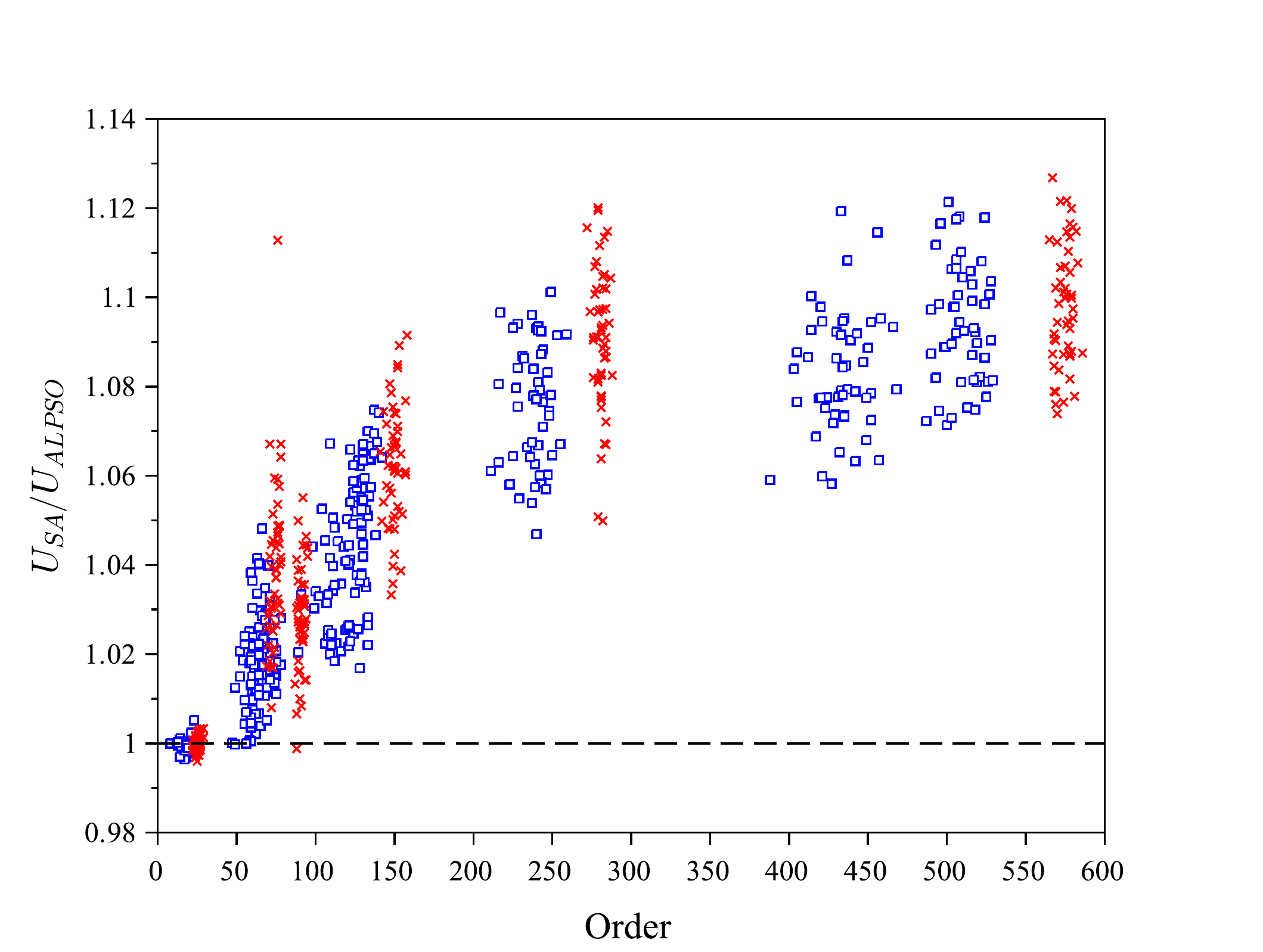}}}\hspace{2mm}
\subfigure[Average betweenness centrality.]{\adjustbox{trim={.02\width} {.07\height} {0.1\width} {.1\height},clip}  {\includegraphics[width=0.54\textwidth]{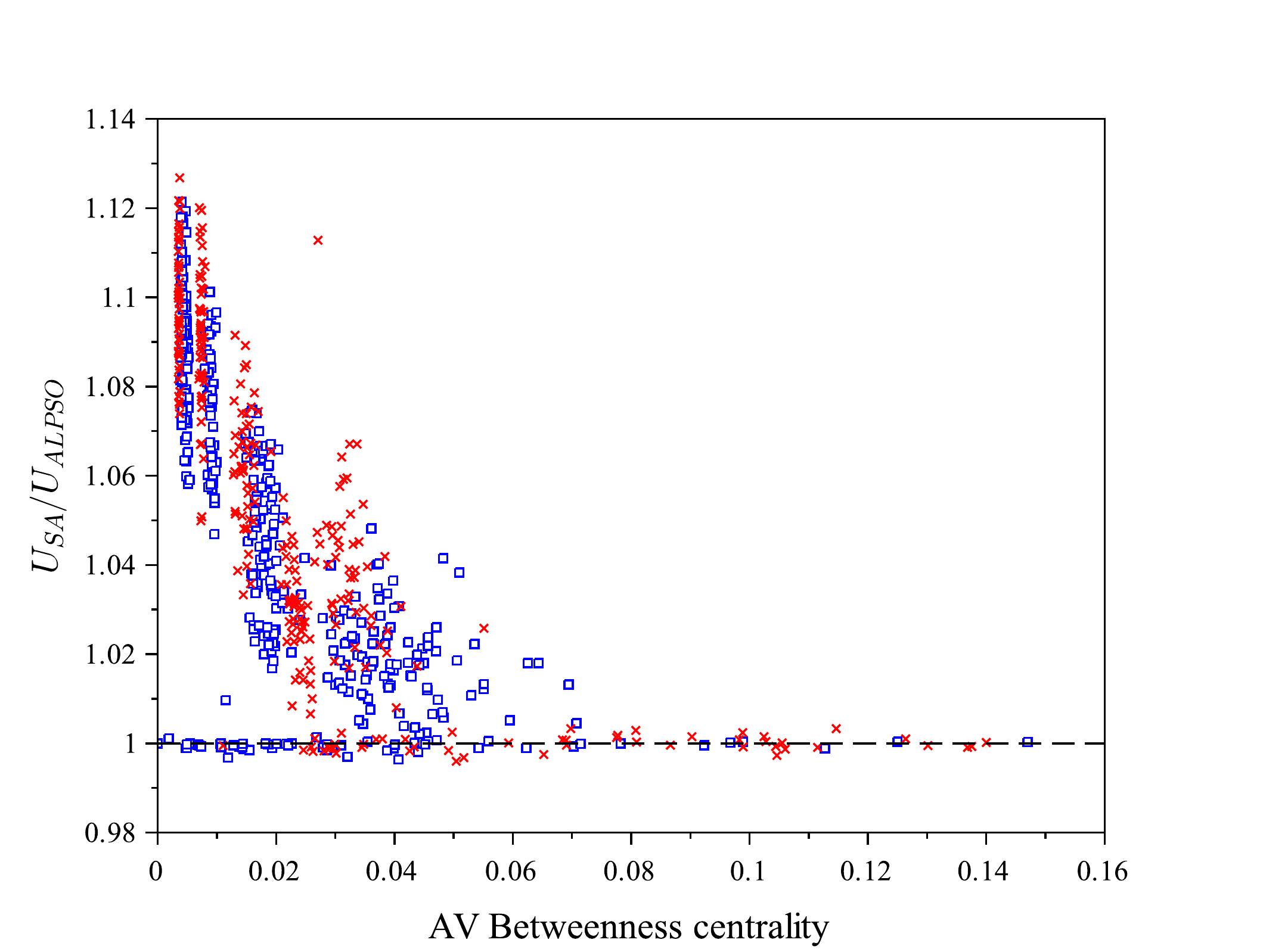}}}\vspace{0mm}
\subfigure[Density.]{\adjustbox{trim={.02\width} {.07\height} {0.1\width} {.1\height},clip}  {\includegraphics[width=0.54\textwidth]{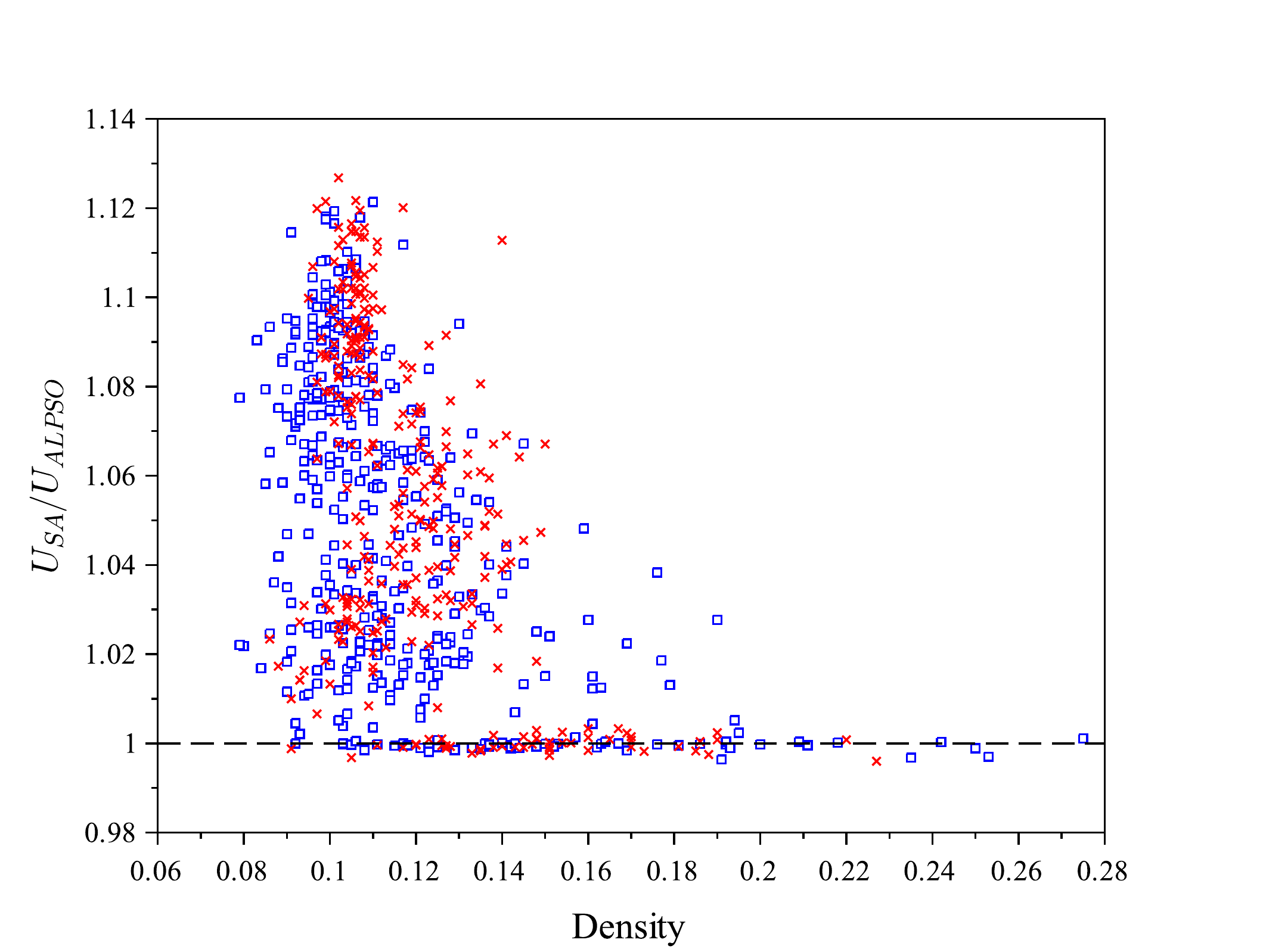}}}\hspace{2mm}
\subfigure[Diameter.]{\adjustbox{trim={.02\width} {.07\height} {0.1\width} {.1\height},clip}  {\includegraphics[width=0.54\textwidth]{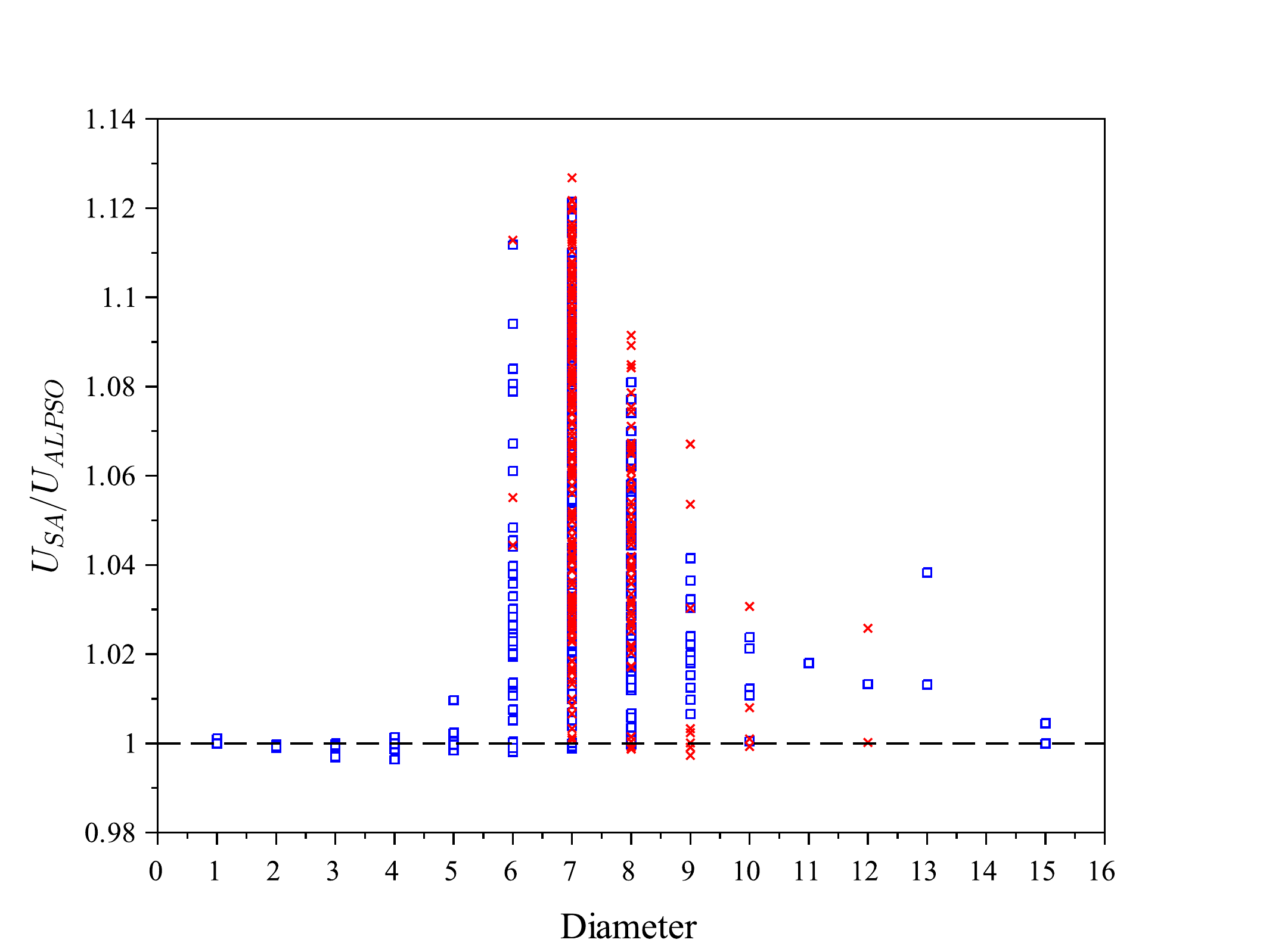}}}
\subfigure[Cluster coefficient.]{\adjustbox{trim={.02\width} {.07\height} {0.1\width} {.1\height},clip}  {\includegraphics[width=0.54\textwidth]{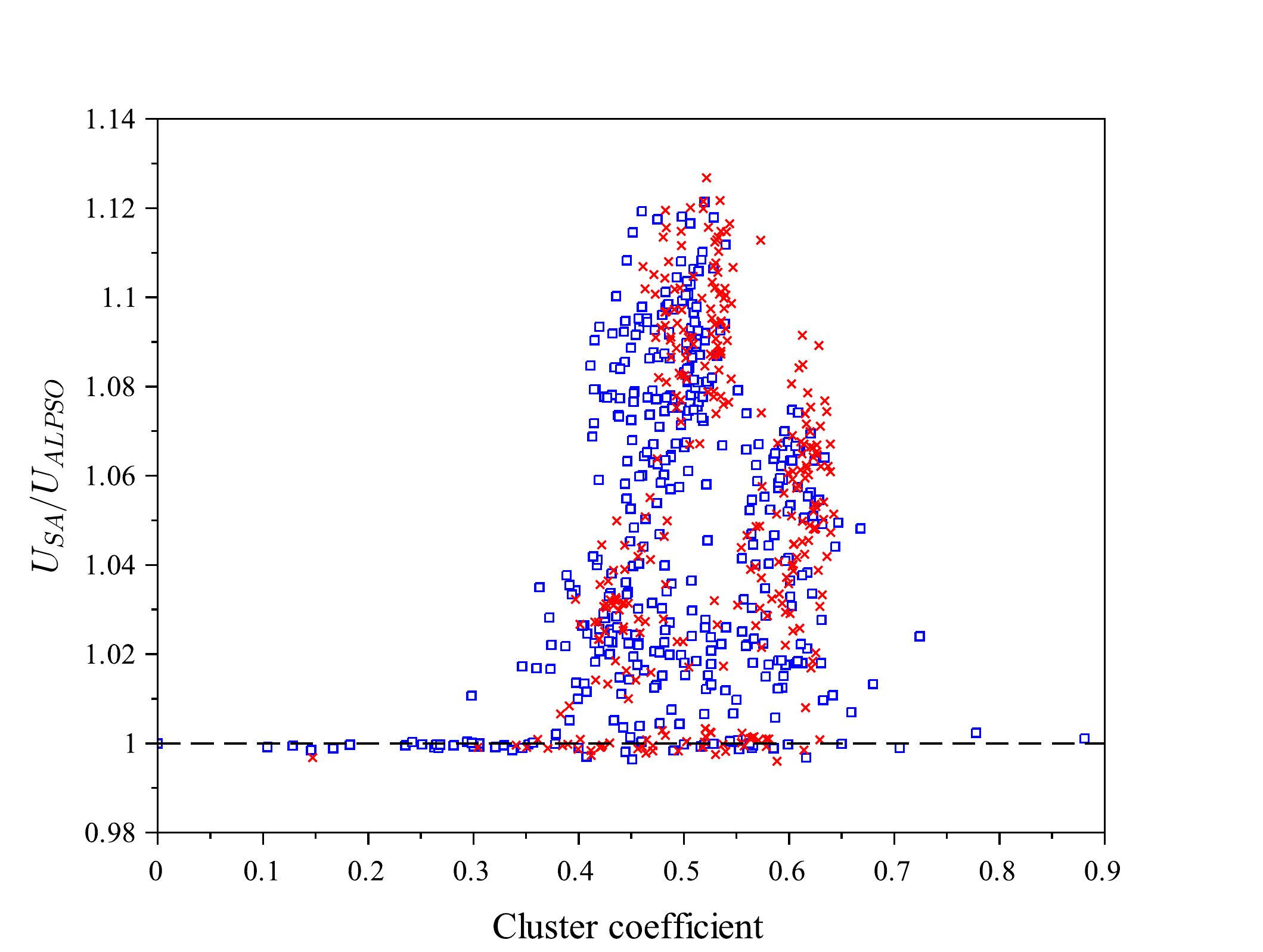}}}\hspace{2mm}
\subfigure[Wiener index.]{\adjustbox{trim={.02\width} {.07\height} {0.1\width} {.1\height},clip}  {\includegraphics[width=0.54\textwidth]{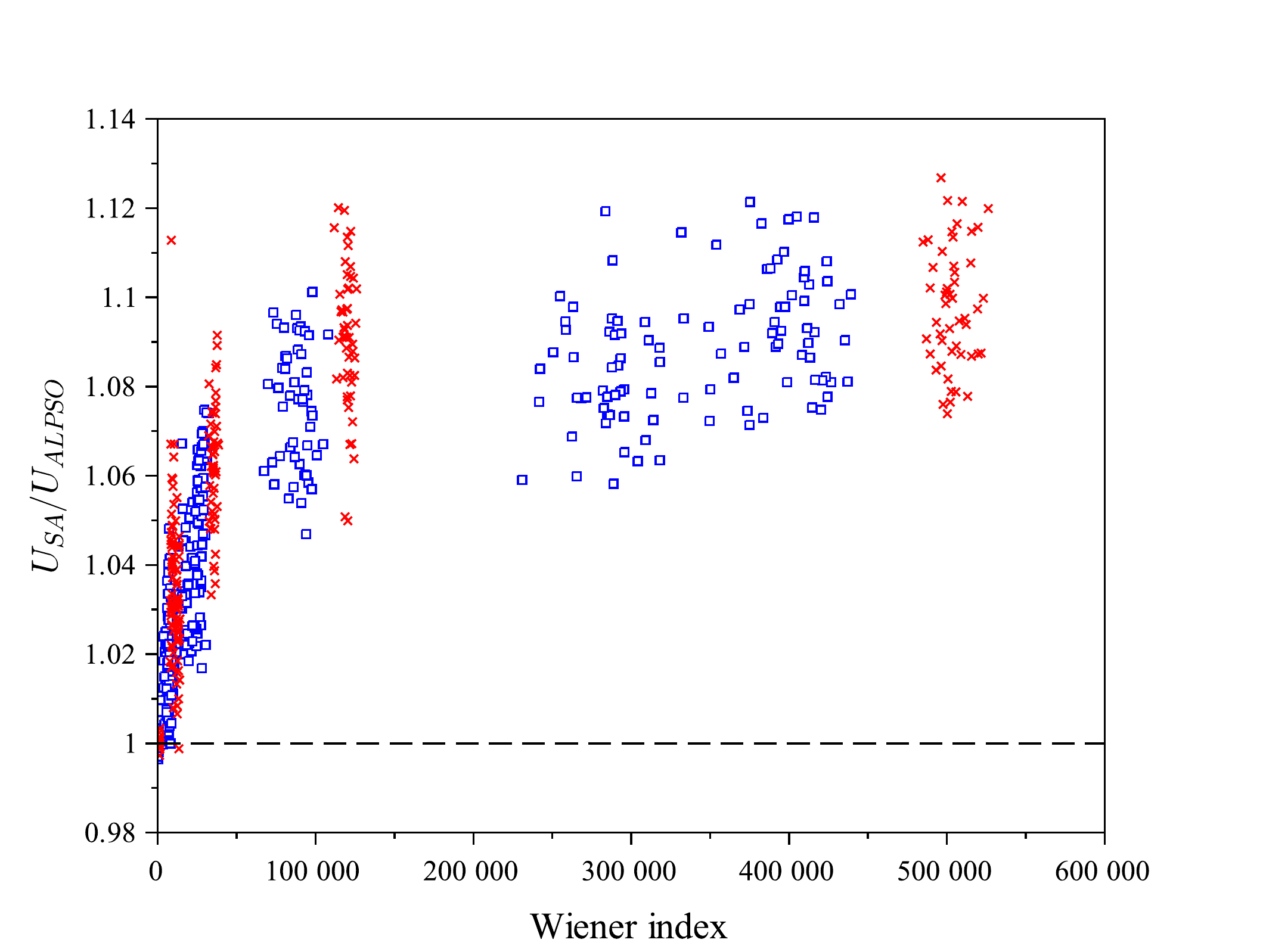}}}
\vspace{-3mm}
\caption{Utility of SA relative to ALPSO for different graph metrics.} \label{fig:comparison}
\end{figure*}



To account for the diversity of scenarios within each category, we have analyzed the results of the best performing approach (SA) against the complete-information reference (ALPSO) with respect to the different metrics discussed in Section \ref{metrics}. Figure \ref{fig:comparison} shows, for each metric, the ratio between the average utility achieved by SA in the 10 runs for a given graph, and the average utility obtained by ALPSO for the same graph (hence the dashed line in the figures corresponds to the ALPSO 1.0 baseline). Note also that blue squares correspond to random scenarios while red crosses correspond to square ones. We can see there is an approximately linear increasing gain for SA with graph order, with ALPSO doing better for low-order graphs and SA getting to gains up to 10\% for the larger graphs (Figure \ref{fig:comparison}a). This is coherent with the results in Tables~\ref{tab:utilities_rnd} and~\ref{tab:utilities_sq}. We can see an inversely proportional trend with the average betweenness centrality (Figure \ref{fig:comparison}b). The SA negotiator performs better for low centrality values, which seems reasonable because in these graphs there will be more peripheral nodes (i.e. with less interfering nodes) than central nodes (i.e. with more interfering nodes), which should make negotiations easier. The same reasoning explains the results with respect to density (Figure \ref{fig:comparison}c). The negotiator fares better in the less dense graphs (i.e. where there are less interference links).

There are other interesting patterns arising from the metrics analysis. For instance, Figures \ref{fig:comparison}d and \ref{fig:comparison}e suggest that there may be optimal values of graph diameter and graph cluster coefficient, respectively, regarding the performance of the SA negotiator. However, further analysis would be needed to rule out other possible explanations. For instance, it is reasonable to expect very little room for improvement of the negotiator in the extremely high clustering coefficient cases (almost complete graph, all nodes interfere with each other).

\section{Conclusions and future work}
\label{conclusiones}

In this work we study the problem of coordinating frequency assignment for Wi-Fi access points. We consider an approach inspired in the well-known graph coloring problem. In contrast with the traditional viewpoint from discrete optimization, we provide a negotiation approach based on a simple-text mediation protocol and agent strategies based on simulated annealing. After experimental evaluation, our results are significantly better than the reference approaches. This is specially significant, because 1) it is the first time (to our knowledge) that nonlinear negotiation is used for  complex network optimization in a realistic setting, and 2) the approach effectively addresses complex negotiation scalability, which has been a challenge to apply this kind of techniques to real settings.

There are a number of future lines of research which emerge from this work. We are studying the creation of negotiation-based hyper-heuristics, to dynamically adapt the approaches to the metrics that characterize the scenarios. We are also working on distributed belief propagation as a way to conduct the negotiation, which raises some challenges in itself. Finally, we are enlarging the family of negotiation approaches under evaluation, and we are working on generalizing the approach to other network-structured real-world problems.

\section{Acknowledgements}
This work has been supported by the Spanish Ministry of Economy and Competitiveness
Grants TIN2016-80622-P (AEI/FEDER, UE), TIN2014-61627-EXP, MTM2017-83750-P.



\printbibliography

\end{document}